\documentclass[reprint,
    preprintnumbers,
    superscriptaddress,
    nofootinbib,
     amsmath,amssymb,
     aps,
     prd,
    floatfix,
longbibliography
]{revtex4-2}

\usepackage{verbatim}
\usepackage{graphicx}
\usepackage{dcolumn}
\usepackage{bm}
\usepackage{xcolor}
\usepackage{pifont}
\usepackage{hyperref}
\usepackage{stmaryrd}
\usepackage{ragged2e}
\hypersetup{
  colorlinks   = true, 
  urlcolor     = blue, 
  linkcolor    = blue, 
  citecolor   = red 
}
\usepackage{caption}
\usepackage{subcaption}
\usepackage{lipsum}
\usepackage{bbm}
\usepackage{textcomp}

\newcommand{\new}[1]{#1}

\begin{document}

\title{Stochastic Inflation in Numerical Relativity}

\author{Yoann L. Launay}
 
  \email{yoann.launay@outlook.com}
 \affiliation{Centre for Theoretical Cosmology, Department of Applied Mathematics and Theoretical Physics,
University of Cambridge, Wilberforce Road, Cambridge CB3 0WA, United Kingdom}

\author{Gerasimos I. Rigopoulos}
\email{gerasimos.rigopoulos@ncl.ac.uk}
\affiliation{School of Mathematics, Statistics and Physics,  Newcastle University, Newcastle upon Tyne, NE1 7RU, United Kingdom}

\author{E. Paul S. Shellard}
\email{eps1@cam.ac.uk}
\affiliation{Centre for Theoretical Cosmology, Department of Applied Mathematics and Theoretical Physics,
University of Cambridge, Wilberforce Road, Cambridge CB3 0WA, United Kingdom}

\begin{abstract}
{
A set of 3+1 equations for stochastic inflation incorporating all metric and scalar matter degrees of freedom, first presented in previous work \cite{Launay24}, are re-derived in a gauge invariant manner. We then present numerical implementations of these stochastic equations, cast in the BSSN formulation of Numerical Relativity, demonstrating their efficacy in both a slow-roll and an ultra slow-roll scenario. We find the evolution is correctly reproduced for all the dynamical variables, and the energy and momentum constraints are well-satisfied. This demonstrates that the stochastic equations are theoretically and numerically robust and ready to be applied to a wider inflationary landscape. Our simulations result in real space realizations of the fully non-linear stochastic dynamics with \new{gradients and anisotropic expansion retained. This work generalizes standard stochastic inflation, inflationary numerical relativity and lattice cosmology, opening up the possibility for reliable predictions of non-perturbative phenomena and providing} precise initial conditions for subsequent cosmological eras. 
}

\end{abstract}
\maketitle
\section{Introduction \label{sec:intro}}
Despite extensive theoretical advances, inflation remains relatively poorly constrained observationally. Cosmic microwave background (CMB) or large-scale structure (LSS) data for instance \cite{PlanckNG, Chaussidon25,Chudaykin25}, is still too limited to probe inflationary interactions, the imprints of which \new{start manifesting} in the so-called 3-point statistics. However, while ongoing experiments promise enriched data in the coming years, in particular on non-linear and potentially non-perturbative scales, substantial challenges remain when modeling connections from high-energy theories to data, and vice-versa. In fact, many scenarios of inflation remain untestable because current computational methods, whether theoretical or numerical, rely on numerous assumptions. For instance, quantum field theory on curved spacetimes \cite{WaldQFTCS, Maldacena_2003, mulryne_pytransport_2017,werth23, Costantini25} is usually limited to tree-level computations and thus truncated and perturbative non-linearity. On the other hand, \new{commonly used} approaches enabling non-perturbative and non-linear studies --- \new{which include stochastic inflation \cite{Starobinsky_stochastic_1988, morikawa_dissipation_1990}, stochastic and standard $\Delta {\cal N}$ formalisms \cite{wands_new_2000, vennin_correlation_2015},  Hamilton-Jacobi approaches \cite{salopek_stochastic_1991, prokopec_ensuremathdeltan_2021}, lattice cosmology \cite{latticeeasy, Cosmolattice, Caravano22thesis}, and numerical relativity for inhomogeneous inflation \cite{East16,clough_robustness_2017,Bloomfield19,joana_inhomogeneous_2021, aurrekoetxea_effects_2020, Corman23,Elley2024, Ijjas22}} ---do not fully complement QFT as they rely on gradient expansions, or approximate relativistic and quantum-to-classical effects. 
Beyond standard types of inflation, the landscape of testable scenarios is thus limited to those where strong leading-order approximations are valid.

With the expected arrival of smaller scales probes, Stochastic Inflation has become a popular non-perturbative method for studying the non-linear semi-classical evolution of long-wavelength inhomogeneities during inflation, while still accounting for quantum diffusion. Its standard formulation relies on the separate universe approximation (SUA) or $0th$-order gradient expansion, formulated in a specific gauge choice with the scalar field as the fluctuating dynamical variable. The so-called stochastic $\Delta {\cal N}$ formalism is then used to translate field fluctuations into statistics for the curvature perturbation \cite{Starobinsky_stochastic_1988, Finelli09, Finelli10, pattison_stochastic_2019,Jackson22,figueroa_non-gaussian_2021,Figueroa22,tomberg_numerical_2023,Mizuguchi24}. \new{Recent work within the existing stochastic inflation formalism and $\Delta\mathcal{N}$ \cite{artigas_hamiltonian_2022, jackson2023,Tanaka:2023gul,Tanaka:2024mzw, Briaud25} has aimed to improve on the use of  gradient expansions, gauge artifacts and the neglect of anisotropic perturbations but, overall, stochastic inflation remains firmly rooted on the SUA.} 

\new{Our recent work \cite{Launay24} showed that the SUA, $\Delta {\cal N}$ and specific gauge choices are not required to formulate stochastic inflation by deriving a direct generalization of the stochastic equations within full General Relativity. The original derivations in \cite{Launay24}, involved specific gauge families. In this work we re-derive these equations without picking a gauge. These stochastic equations include all scalar and tensor degrees of freedom, metric and field alike, and are valid for any choice of time-slicing and threading of the spacetime. The derivation involves coarse-graining on  hypersurfaces by expressing all variables on them in terms of $\mathcal{R}$, the coarse-grained curvature perturbation on comoving hypersurfaces. We then present a Numerical Relativity scheme where these stochastic inflation equations are solved for scalar sources in the synchronous gauge for two test scenarios: a slow roll potential and a potential with an inflection point with a transition to ultra slow-roll. This is distinct from our previous investigation of these scenarios with full GR evolution \cite{Launay2025}, in that all perturbations were part of the initial conditions there (see also \cite{Florio2024} for tensor modes). The current work fully incorporates stochastic noise terms that act as modes continuously cross the coarse-graining scale.}

We present the ADM stochastic equations and outline their derivation in Section \ref{sec:Stoch-Eqns}. Section \ref{sec:istoriz} introduces the numerical implementation, and validates on the two aforementioned applications, including a first look at anisotropic (shear)  DOFs. We conclude in Section \ref{sec:Conclusion} and leave in the Appendices the derivation of scalar gauge invariants in terms of ${\cal R}$ (\ref{app:comProof}), the full set of stochastic BSSN equations (\ref{app:BSSN}) and details of the discretization schemes (\ref{app:schemes} and \ref{app:theta}). 

\section{Stochastic General Relativity}\label{sec:Stoch-Eqns}
\new{In this section we derive the general relativistic equations for stochastic inflation in a gauge invariant manner. We begin by briefly recalling the standard, phase space formulation.}

\new{\subsection{Stochastic Inflation primer}}

\new{In his original derivation of stochastic inflation, A. Starobinsky \cite{Starobinsky_stochastic_1988} showed that field fluctuations during inflation can be separated depending on their wavelength compared to the Hubble radius $(aH)^{-1}$. Beyond the latter scale, modes (noted here with superscript ${}^>$) behave classically and can be accounted for in a classical IR theory.
Given that the Hubble radius shrinks in comoving coordinates, modes become progressively part of the IR system of fluctuations. This results in random kicks in the associated equations of motion (EOM), effectively stochastic PDEs, which can then be evolved completely nonlinearly and nonperturbatively-- as opposed to the UV quantum theory. }

\new{
Over the past decades, Stochastic Inflation has undergone-- and still sees-- extensive validity studies. A popular formulation is that summarised in \cite{pattison_stochastic_2019}, formulated in the uniform efolds gauge in general relativity, and on scales much larger than the Hubble radius, where patches of the universe are then evolving independently according to the following Langevin equations
\begin{equation}
\left \{
\begin{aligned}
 &\frac{\partial \phi^>}{\partial{\cal N}_b} -\pi^> = \boldsymbol{\Sigma_{\phi}}, \\
 &\frac{\partial \pi^>}{\partial{\cal N}_b}   + (3-\varepsilon_1^>)\pi^>+\frac{1}{(H^>)^2}\frac{dV}{d\phi}(\phi^>)= \boldsymbol{\Sigma_{\pi}}, \\
& {H^>}^2  = \frac{V(\phi^>)}{1-\frac{1}{6M_{Pl}^2}(\pi^>)^2},
\end{aligned}\right .
    \label{eq:SIeq}
\end{equation}
where ${\cal N}_b=\ln a$ is the background number of e-folds, $\varepsilon_1^>=-\partial_{{\cal N}_b}\ln H^>$ the first slow-roll IR parameter, and where the random fields are defined from Fourier space
\begin{equation}
\left \{
\begin{aligned}
\boldsymbol{\Sigma_{\phi}} & = {\cal F}^{-1}\left\{\frac{\partial W}{\partial{\cal N}_b}\delta \phi_{\mathbf{k}}\boldsymbol{\alpha_{\mathbf{k}}}\right\} , \\
\boldsymbol{\Sigma_{\pi}} & = {\cal F}^{-1}\left\{ \frac{\partial W}{\partial{\cal N}_b}\delta \pi_{\mathbf{k}}\boldsymbol{\alpha_{\mathbf{k}}}\right\}.
\end{aligned}\right .
    \label{eq:SInoise}
\end{equation}
The statistics are that of gaussian random variables $\boldsymbol{\alpha_{\mathbf{k}}}$ such that for all $(\boldsymbol{\mathbf{k}} , \boldsymbol{\mathbf{k}'})$ in the same Fourier hemisphere
\begin{equation}
\left\{
\begin{aligned}
         \langle \boldsymbol{\alpha_{\mathbf{k}}}\boldsymbol{\alpha_{\mathbf{k}'}} \rangle_{\mathbb{P}}  & = 0, \\
           \langle \boldsymbol{\alpha_{\mathbf{k}}}\boldsymbol{\alpha_{\mathbf{k}'}}^* \rangle_{\mathbb{P}} & \propto  \delta^{(3)}(\mathbf{k}-\mathbf{k}').
\end{aligned}\right . \label{eq:alphas}
\end{equation}
}
\new{
As noted in various approaches for extension, including recent works \cite{cruces_review_2022, cruces_stochastic_2022,  artigas_hamiltonian_2022,jackson2023, Briaud25}, these equations suffer from a consequent number of approximations (gauge and constraint subtleties, neglecting Hubble scale interactions, no gradient couplings between neighbouring points), subject to breaking in scenarios and predictions of interest. }

\new{
In \cite{Launay24}, we proposed a way forward to solve these issues. The associated stochastic equations are recalled in \eqref{eq:CoarseGrainedADM_Dyn3}, \eqref{eq:CoarseGrainedADM_Dyn1}, \eqref{eq:CoarseGrainedADM_Dyn2} 
and
\eqref{eq:CoarseGrainedADM_Con1}, \eqref{eq:CoarseGrainedADM_Con2}. In that work the source terms were obtained for diverse gauge families and coincided for all of them, leading us to conjecture their gauge independence. In this section we provide a rigorous proof that these stochastic equations are indeed valid in all gauges. 
}
\subsection{Coarse-graining using $\mathcal{R}$}

The derivation of the stochastic equations starts by considering a perturbed inflationary spacetime with a metric
\begin{equation}
\begin{aligned}
    d s^2=-\alpha_b^2(1+2 \Psi) d t^2+2 a^2 B_{, i} d t d x^i & \\
    +a^2\left[(1-2 \Phi) \delta_{i j}+2 E_{, i j}\right] d x^i d x^j,
\end{aligned}    
    \label{eq:SVTmetric}
\end{equation}
where the four functions $\Phi, \Psi, B$ and $E$ define the metric's scalar perturbations, with $\alpha_b$ and $a$ the background lapse and scale factor. The scalar field driving the expansion is also perturbed as $\phi =\phi_b + \delta \phi$. The perturbations can be combined in the well-known linear gauge invariant variables
\begin{equation}
    \left \{\begin{array}{cl}
         \Phi + H\chi & \equiv\Phi_B , \\
         \Psi - \dot{\chi} & \equiv \Psi_B ,\\
        \delta\phi-{\dot{\phi_b}}\chi  & \equiv \delta\phi_{gi},\\
          -\Phi+ \frac{\delta\rho}{3(\rho_b+ P_b)}  &\equiv \zeta_{gi}, \\
        \Phi + \frac{H}{\dot{\phi_b}}\delta \phi & \equiv {\cal R} ,
    \end{array}\right .
    \label{eq:GIquantities}
\end{equation}
where $\delta\rho = \alpha_b^{-2}{\dot{\phi_b}}{\delta\dot{\phi}}+{V'}|_{b}\delta\phi-\alpha_b^{-2}\dot{\phi_b}^2\Psi $ is the scalar field energy density perturbation and $ \chi \equiv-{\alpha_b}^{-1}{a^2}(B-\dot{E})$. There are therefore five scalar perturbation variables which can be expressed in terms of the five gauge invariant variables.   

We will take ${\cal R}$ to be the master dynamical perturbation variable via which all others can be obtained. As shown in Appendix~\ref{app:comProof} and without relying on any further assumptions, the constraints equations can be used to obtain all other gauge-invariant variables in terms of $\mathcal{R}$ and its derivatives in Fourier space \cite{Launay24}  
    \begin{equation}
    \left\{\begin{array}{cl}
         \Phi_{B}  & = -\varepsilon_1 H a^2 k^{-2}\dot{{\cal R}}, \\
         \Psi_{B}  & = \varepsilon_1{\cal R}  + \varepsilon_1a^2k^{-2}\left[\ddot{{\cal R}} +H(2-\varepsilon_2)\dot{{\cal R}} \right],\\
         \delta\phi_{gi}  & =  \sqrt{2\varepsilon_1}{\rm M_{Pl}}\left[{\cal R} +\varepsilon_1a^2 H k^{-2}\dot{{\cal R}} \right],\\
         \zeta_{gi}  &= -{\cal R} +\frac{1}{3} H^{-1}\dot{{\cal R}}, \\
    \end{array}\right .
    \label{eq:GaugeInvExplicit}
\end{equation}
where the slow-roll parameters are $ \varepsilon_{i+1} \equiv -H^{-1}d_t \ln \varepsilon_i$, $\varepsilon_0 \equiv H$, in particular  $\dot{\phi}_b^2 = 2\varepsilon_1H^2{\rm M_{Pl}^2}$.

The first step in obtaining our stochastic equations is to now \emph{coarse grain \new{the Fourier amplitude of} the fundamental variable} ${\cal R}$
\begin{equation}\label{eq:replacement}
{\cal R}_k\longrightarrow {\cal R}_k^> = W_k {\cal R}_k\,,
\end{equation}
where $W_k$ is a window function chosen to have support on long wavelengths: $W_0=1$ and $\lim_{k\to \infty} W_k = 0$, with a relatively steep drop around a value $k=\sigma aH$ where $\sigma$ a parameter determining the coarse-graining scale in terms of the Hubble radius $R_H=(aH)^{-1}$. The step function $W_k=\Theta(\sigma aH-k)$ is a common choice but smoother functions can also be envisaged. Here, we will use the smooth window \eqref{eq:ourWindow}.

From now on coarse grained time derivatives are to be understood as $\dot{f}^> \equiv d\left(Wf\right)/dt$
in any linearized perturbation variable $f$. 
From the Sasaki-Mukhanov equation
\begin{equation}
\ddot{{\cal R}}_k + H(3-\varepsilon_2)\dot{{\cal R}}_k+ \frac{k^2}{a^2}{\cal R}_k = 0,
    \label{eq:Req}
\end{equation}
we then find that the coarse grained variable ${\cal R}_k^>$ obeys
\begin{equation}
\ddot{{\cal R}}^{>}_k + H(3-\varepsilon_2)\dot{{\cal R}}^{>}_k+ \frac{k^2}{a^2}{\cal R}^{>}_k = {\cal S}_{{\cal R}},
    \label{eq:coarseGrainedReq}
\end{equation}
where the spectral source term appearing on the r.h.s. is
\begin{equation}\label{eq:SR}
    {\cal S}_{{\cal R}}  \equiv {\cal R}_k \ddot{W}_k+[2 \dot{\cal R}_k + (3-\varepsilon_2) H {\cal R}_k]\dot{W}_k.
\end{equation} 
 Note that ${\cal S}_{{\cal R}}$ vanishes when the window is constant and therefore its existence is solely attributed to the time-dependent UV-IR split or to initial conditions.

The second step is to replace ${\cal R}  \rightarrow {\cal R}^>$ on the r.h.s. of \eqref{eq:GaugeInvExplicit} and use \eqref{eq:coarseGrainedReq} to replace the second time derivative term. This  \emph{defines} our coarse-grained linear gauge invariant variables as 
\begin{equation}
    \left\{\begin{array}{cl}
         \Phi_{B}^> & =-\varepsilon_1a^2k^{-2}H\dot{{\cal R}}^> ,\\
         \Psi_{B}^>  & =  -\varepsilon_1a^2k^{-2}(H\dot{\cal R}^>-{\cal S}_{\cal R}),\\
         \delta\phi_{gi}^>  & =  \sqrt{2\varepsilon_1}{\rm M_{Pl}}\left[{\cal R}^> +\varepsilon_1a^2 H k^{-2}\dot{{\cal R}}^> \right],\\
         \zeta_{gi}^>  &= -{\cal R}^> +\frac{1}{3} H^{-1}\dot{{\cal R}}^>. \\
    \end{array}\right .
    \label{eq:GaugeInvExplicit>}
\end{equation}
These relations are then inserted into the r.h.s. of \eqref{eq:GIquantities} to define the coarse grained scalar perturbation variables $\Phi^>$, $\Psi^>$, $\chi^>$ and $\delta\phi^>$ in terms of ${\cal R}^>$ and $\dot{{\cal R}}^>$, also explicitly containing ${\cal S}_{\cal R}$
\begin{equation}
    \left \{\begin{array}{cl}
         \Phi^> + H\chi^> & = -\varepsilon_1a^2k^{-2}H\dot{{\cal R}}^>  , \\
         \Psi^> - \dot{\chi}^> & =  -\varepsilon_1a^2k^{-2}(H\dot{\cal R}^>-{\cal S}_{\cal R}),\\
        \delta\phi^>-{\dot{\phi_b}}\chi^>  & = \sqrt{2\varepsilon_1}{\rm M_{Pl}}\left[{\cal R}^> +\varepsilon_1a^2 H k^{-2}\dot{{\cal R}}^> \right],\\
          -\Phi^>+ \frac{\delta\rho^>}{3(\rho_b+ P_b)}  &=  -{\cal R}^> +\frac{1}{3} H^{-1}\dot{{\cal R}}^>, \\
        \Phi^> + \frac{H}{\dot{\phi_b}}\delta \phi^> & = {\cal R}^> .
    \end{array}\right .
    \label{eq:CGvariables}
\end{equation}
Each of the coarse-grained scalar perturbation variables of the l.h.s. can now be solved for, \new{(i.e. expressed in terms of ${\cal R}^>$ and  $\dot{{\cal R}}^>$)} and as a final step inserted into the linearized version of Einstein's equations in their ADM form. After further use of \eqref{eq:coarseGrainedReq} and resulting cancellations of terms, the only terms that survive involve ${\cal S}_{\cal R}$, becoming the sources in the stochastic equations. We illustrate this procedure in the following subsection by explicitly performing it for the field equation.

\subsection{Gauge invariant coarse-graining: field equation } \label{app:CGfield}

As an explicit example of the procedure described in the previous subsection, we now demonstrate the coarse-graining of the ADM inflaton equation 
\begin{equation}
 \frac{1}{\alpha}\left(\dot{\Pi}+\beta^i \Pi_{\mid i}\right)-K \Pi-\frac{\alpha^{\mid i}}{\alpha} \phi_{\mid i} 
 -\phi_{\mid i}^{\mid i}+\frac{d V}{d \phi}=0\,,
\end{equation}
where \new{$\alpha$ s the shift, $\beta^i$ the lapse}, $\Pi=\alpha^{-1}(\dot{\phi}+\beta^i \partial_i \phi)$ and, as usual, $K=-3H$ measures the local volume expansion - see \ref{subsec:FASI} for the definition of ADM variables.
Before coarse-graining, the corresponding linear field equation in Fourier space is
\begin{equation}
\begin{aligned}
   \delta\ddot{\phi}+3 H  \delta \dot{\phi}+\frac{k^2}{a^2} \delta \phi
    -\dot{\phi_b}\left(3\dot{\Phi}+\dot{\Psi}+\frac{k^2}{a^2}\chi\right)\\
   +\frac{\mathrm{d}^2V}{\mathrm{d}\phi^2}(\phi_b) \delta \phi +2 \Psi\frac{\mathrm{d}V}{\mathrm{d}\phi}(\phi_b)=0
    \end{aligned}
        \label{eq:fieldCPT}
\end{equation}
where $\alpha_b$ was set to $1$ without loss of generality. We want to calculate the spectrum of the $r.h.s.$ of the same equation but for the IR field perturbations $(\delta\phi^>, \Phi^>, \Psi^>, \chi^>)$.

Using eq. \eqref{eq:CGvariables} and \emph{without picking a gauge}, it is possible to express all these linear degrees of freedom as functionals of ${\cal R}^>$ and another scalar of our choice. In the case of the field equation, we choose this scalar to be the inflaton's perturbation $\delta\phi^>$. Eq. \eqref{eq:CGvariables} can then be rearranged
\begin{equation}
    \left \{\begin{array}{cl}
        \Phi^>  & = {\cal R}^>-{\dot{\phi_b}}^{-1}{H}\delta \phi^>\\
        \chi^>  & = -H^{-1}\Phi^> -\varepsilon_1a^2 k^{-2}\dot{{\cal R}}^> ,\\
         \Psi^>  & =\dot{\chi}^{\rm >}  -\varepsilon_1a^2k^{-2}(H\dot{\cal R}^>-{\cal S}_{\cal R}).
    \end{array}\right .
    \label{eq:CGvariables1}
\end{equation}
By making further use of eq. \eqref{eq:coarseGrainedReq}, the time differentiations needed in the coarse-graining of \eqref{eq:fieldCPT} can be expressed as 
\begin{equation} 
    \left \{\begin{array}{cl}
         \dot{\Phi}^> & = \dot{\cal R}^> - H\dot{\phi}_b^{-1}{\delta \dot{\phi}^>}-\frac{1}{2}\varepsilon_2H^2\dot{\phi}_b^{-1}{\delta \phi^>} \\[0.4em]
         
         \dot{\chi}^>  & = H(\varepsilon_1+\frac{1}{2}\varepsilon_2)\dot{\phi}_b^{-1}{\delta\phi^>}+\dot{\phi}_b^{-1}{\delta \dot{\phi}^>}-H^{-1}\dot{\cal R}^> \\[0.4em]
         & +\varepsilon_1a^2k^{-2}(H\dot{\cal R}^>-{\cal S}_{\cal R}) \\[0.4em]
        
         \dot{\Psi}^> & =  \dot{\phi}_b^{-1}{\delta\ddot{\phi}^>}+(2\varepsilon_1+\varepsilon_2)H\dot{\phi}_b^{-1}\delta\dot{\phi}^>\\[0.4em]
         & +\frac{1}{4}\varepsilon_2(-2\varepsilon_1+\varepsilon_2-2\varepsilon_3)H^2\dot{\phi}_b^{-1}\delta\phi^>\\[0.4em]
         &+a^2H^{-1}k^2{\cal R}^>+(3-\varepsilon_1-\varepsilon_2)\dot{\cal R}^>\\[0.4em]
         & -H^{-1}{\cal S}_{\cal R}.
    \end{array}\right .
    \label{eq:CGvariables2}
\end{equation}
Replacing these, together with the previous eq. \eqref{eq:CGvariables1} and
\begin{equation}
\left \{
\begin{array}{ll}
\displaystyle\frac{\mathrm{d}V}{\mathrm{d}\phi}(\phi_b) & = (-3+\varepsilon_1+\varepsilon_2/2) H \dot{\phi}_b^{-1}, \\[0.7em]
\displaystyle\frac{\mathrm{d}^2V}{\mathrm{d}\phi^2}(\phi_b) & = -\frac{1}{4}[8\varepsilon_1^2+2\varepsilon_1(-12+5\varepsilon_2)\\[0.4em]
&+\varepsilon_2(-6+\varepsilon_2+2\varepsilon_3)] H^2, \\
\end{array}\right .
\end{equation}
into the the l.h.s. of \eqref{eq:fieldCPT}, all terms perfectly cancel apart from the Mukhanov-Sasaki-Starobinsky term ${\cal S}_{\cal R}$ which survives as a source term. One therefore sees that the IR version of eq. \eqref{eq:fieldCPT} is
\begin{equation}
\begin{aligned}
   \delta\ddot{\phi}^>+3 H  \delta \dot{\phi}^>+\frac{k^2}{a^2} \delta \phi^>
    -\dot{\phi_b}\left(3\dot{\Phi}^>+\dot{\Psi}^>+\frac{k^2}{a^2}\chi^>\right)\\
    +\frac{\mathrm{d}^2V}{\mathrm{d}\phi^2}(\phi_b) \delta \phi^> +2 \Psi^>\frac{\mathrm{d}V}{\mathrm{d}\phi}(\phi_b)= \dot{\phi}_bH^{-1} {\cal S}_{\cal R}\,,
    \end{aligned}
        \label{eq:fieldCPT-CG}
\end{equation}
in agreement with the equation presented in previous work \cite{Launay24}, but proven here directly \emph{for all gauge choices}. 

Similarly to standard derivations of stochastic inflation, the obtained r.h.s. is now assumed to provide the spectrum of the random variable making up the stochastic noise term for the full non-linear version of \eqref{eq:fieldCPT}, giving the {stochastic} ADM inflaton equation
\begin{equation}
 \frac{1}{\boldsymbol{\alpha}}
 \left(
   \dot{\boldsymbol{\Pi}}
   + \boldsymbol{\beta}^i \boldsymbol{\Pi}_{\mid i}
 \right)
 - \boldsymbol{K}\boldsymbol{\Pi}
 - \frac{\boldsymbol{\alpha}^{\mid i}}{\boldsymbol{\alpha}} \boldsymbol{\phi}_{\mid i}
 - \boldsymbol{\phi}_{\mid i}^{\mid i}
 + \frac{\mathrm{d}V}{\mathrm{d}\phi}(\boldsymbol{\phi})
 = \mathcal{F}^{-1}\{\boldsymbol{\mathcal{S}}_{\Pi}\},
 \label{eq:CoarseGrainedADM_Dyn3}
\end{equation}
where ${\cal F}^{-1}\{\cdot\}$ is the inverse hermitian Fourier transform \new{and where 
\begin{equation}
    \boldsymbol{{\cal S}}_{\Pi} ={\rm M_{Pl}} \sqrt{2\varepsilon_1}   \boldsymbol{{\cal S}}_{{\cal R}}. 
    \label{eq:SPi}
\end{equation}
Here, we have defined the stochastic noise source with the spectrum given in eq.~\eqref{eq:SR}
\begin{equation}
\boldsymbol{\mathcal{S}}_{\mathcal{R}} (\mathbf{k})= \boldsymbol{{\cal R}}_{\mathbf{k}} \ddot{W}_k+[2 \dot{\boldsymbol{{\cal R}}}_{\mathbf{k}} + (3-\varepsilon_2) H \boldsymbol{{\cal R}}_{\mathbf{k}}]\dot{W}_k.
\end{equation}
where
\begin{equation}
   \boldsymbol{{\cal R}}_{\mathbf{k}}(t) =   {\cal R}_k(t){}\boldsymbol{\alpha_{\mathbf{k}}}
\end{equation}   
and the statistics the stochastic Gaussian basis $\{\boldsymbol{\alpha}\}$ is given by  equation~\eqref{eq:alphas}.
In practice, the random variables $\boldsymbol{\alpha_{\mathbf{k}}}$ are drawn once at some fixed early time and that defines a real space random field $\boldsymbol{{\cal R}}(t,\mathbf{x})$ that will be propagated in time by the real space version of \eqref{eq:Req}, written as
\begin{equation}
\left\{
    \begin{aligned}
        \boldsymbol{\Pi}_{\cal R} &= \frac{9\dot{\phi}_b^2}{\alpha_b K_b^2} \bigg(\partial_0\boldsymbol{{\cal R}}+\beta^i\partial_i\boldsymbol{{\cal R}}\bigg)\\
        \partial_0\boldsymbol{\Pi}_{\cal R}&-\alpha_b K_b\boldsymbol{\Pi}_{\cal R}+\beta_b^i\partial_i\boldsymbol{\Pi}_{\cal R} -\frac{9\dot{\phi}_b^2\alpha_b}{ K_b^2}\partial_i\partial^i\boldsymbol{{\cal R}}=0\,. 
    \end{aligned}\right .
    \label{eq:ReqISTORIZ1}
    \end{equation}
}

\new{We note here that a possible heuristic inclusion of the backreaction of the long wavelength inhomogeneous fields on the UV modes used to compute the noise is the promotion of the background parameters in the equation determining the solution for the UV modes into functions of those spatially-dependent long wavelength fields \cite{salopek_stochastic_1991, levasseur_backreaction_2015, figueroa_non-gaussian_2021,Figueroa22}. Following this general approach, in this work we include the possibility of such long-wavelength backreaction on short wavelength modes by allowing for the coefficients in \eqref{eq:ReqISTORIZ1} to take their local values\footnote{To our knowledge, such a scheme was first proposed but not numerically implemented in \cite{salopek_stochastic_1991}. More recently, a similar construction for the mode equations was numerically implemented in \cite{figueroa_non-gaussian_2021,Figueroa22,Tomberg:2024evi}.}: $\phi_b \rightarrow \phi(t,x^i)$, $K_b \rightarrow K(t,x^i)$ and  $\alpha_b\rightarrow \alpha(t,x^i)$, $\beta^i_b\rightarrow \beta^i(t,x^i)$ making the noise sources non-Gaussian.}

\subsection{Full ADM Stochastic Inflation}\label{subsec:FASI}

The procedure of inserting the coarse grained perturbation variables in the l.h.s of the linearized versions of the dynamical equations, and thus obtaining the resulting source term, can also be performed for the other dynamical ADM equations. In this section, we present the result of the application of this procedure to these remaining equations: a set of stochastic dynamical equations for the extrinsic curvature of the spatial slices that are valid for any choice of coordinates, while satisfying the usual Energy and Momentum constraints. By construction, standard linear perturbation theory is recovered for the coarse grained variables when this system is linearized. In particular, one can re-derive \eqref{eq:coarseGrainedReq} by working backwards. Clearly, if no time dependent long-short split is performed $\dot{W}_k=0$, all standard equations are recovered. However, equations  \eqref{eq:CoarseGrainedADM_Dyn3} - \eqref{eq:CoarseGrainedADM_Dyn1} - \eqref{eq:CoarseGrainedADM_Dyn2} 
and
\eqref{eq:CoarseGrainedADM_Con1} - \eqref{eq:CoarseGrainedADM_Con2} are non-linear and can capture any non-linear development once modes have crossed the horizon as set by $W$. 
\new{In the following,  the stochastic IR $\boldsymbol{bold}$ notation of the l.h.s. will be kept implicit to avoid notational clutter. }           
 
Using the ADM 3+1 foliation notations for the metric
\begin{equation}
    d s^2=-\alpha^2 d t^2+\gamma_{i j}\left(d x^i-\beta^i d t\right)\left(d x^j -\beta^j d t\right),
    \label{eq:ADMmetric}
\end{equation}
the equations of the ADM formulation of General Relativity are now extended to a set of dynamical stochastic equations. In particular, the extrinsic curvature {of the spacelike 3D hypersurfaces}, defined as $K_{ij}\equiv -\left(2\alpha\right)^{-1}(\gamma_{i j,0}+\beta_{i|j}+\beta_{j|i})$, is found to obey
\begin{eqnarray}
    \dot{K} + \beta^i K_{, i}+\alpha^{\mid i}{ }_{\mid i}-\alpha\left({ }^{3} R+K^2\right)  \nonumber\\
      - {\rm M_{Pl}^{-2}} \alpha\left(\frac{1}{2} S-\frac{3}{2} \rho\right) =  {\cal F}^{-1}\{\boldsymbol{{\cal S}}_{K}\}, \label{eq:CoarseGrainedADM_Dyn1} 
\end{eqnarray}
and
\begin{eqnarray}
\dot{\tilde{K}}_{ij} + 2\alpha\tilde{K}_{il}{\tilde{K}}^l{ }_j+\beta^k \tilde{K}_{ij \mid k}-2\beta_i{ }^{\mid k} \tilde{K}_{jk}  \nonumber\\
       + \alpha_{\mid i\mid j}
      -\frac{1}{3} \alpha^{\mid k}{ }_{\mid k}\delta_{ij} 
    -\alpha\left({ }^3\tilde{R}_{ij}+\frac{1}{3}K \tilde{K}_{ij}\right)  \nonumber\\
    + {\rm M_{Pl}^{-2}} \alpha \tilde{S}_{ij}={\cal F}^{-1}\{\boldsymbol{{\cal S}}_{\tilde{K}_{ij}}\},
    \label{eq:CoarseGrainedADM_Dyn2} 
\end{eqnarray}
where we wrote $K_{ij}=\tilde{K}_{ij}+\frac{1}{3}g_{ij} K$, expressing it in terms of the traceless $\tilde{K}_{ij}$ and trace $K$ parts, and sourced by the stress tensor components $\rho = \alpha^2 T^{00}$, ${\cal J}_i= \alpha T^0{}_i$, $S_{i j} = T_{ij}$,\new{ or more explicitly 
\begin{equation}
\left\{
\begin{aligned}
\rho & =\frac{1}{2} \Pi^2+\frac{1}{2} \partial_i \phi \partial^i \phi+V(\phi), \\
{\cal { J}}_i &= -\Pi \partial_i\phi,
\end{aligned}\right .
\label{eq:densities}
\end{equation}
with the trace and traceless parts of the stress tensor $S_{i j}$ given by
\begin{equation}
    \left \{
    \begin{aligned}
        S & =\frac{3}{2} \Pi^2-\frac{1}{2} \partial_k \phi \partial^k \phi-3 V(\phi), \\
\tilde{S}_{i j}& =  \frac{1}{2}\left(\partial_i \phi \partial_j \phi-\frac{1}{3} \partial_k \phi \partial^k \phi \,\delta_{i j}\right) .
    \end{aligned} \right .
        \label{eq:SETdef}
\end{equation}        }
A vertical bar denotes a covariant derivative associated with $\gamma_{ij}$ and {${ }^3{R}_{ij}[\gamma_{ij}]$ is the Ricci curvature tensor of the spacelike hypersurfaces, also expressed in terms of a trace $R$ and traceless part ${ }^3\tilde{R}_{ij}$.} The source terms appearing on the r.h.s. of \eqref{eq:CoarseGrainedADM_Dyn1} and \eqref{eq:CoarseGrainedADM_Dyn2} are 
\begin{eqnarray}
 \boldsymbol{{\cal S}}_{K} &=&  -\varepsilon_1\boldsymbol{{\cal S}}_{{\cal R}},\\ 
\boldsymbol{{\cal S}}_{\tilde{K}_{ij}} &=& a^2\varepsilon_1(\frac{1}{3}\delta_{ij}-k^{-2}k_ik_j ) \boldsymbol{{\cal S}}_{{\cal R}},
    \label{eq:RHSADM}
\end{eqnarray}
sharing the same random lattice draw as eq.~\eqref{eq:SPi}. Note that a similar coarse-graining can be performed for the quantised graviton degrees of freedom to give an additional source in the $\tilde{K}_{ij}$ evolution \cite{Launay24} but this work focuses on scalar input.

The dynamical equations are supplemented by the two ADM contraints  
\begin{eqnarray}
{\cal H} \equiv { }^{3} R+\frac{2}{3} K^2-\tilde{K}_{i j} \tilde{K}^{i j}-2 {\rm M_{Pl}^{-2}} \rho = 0,  \label{eq:CoarseGrainedADM_Con1}\\
 {\cal M}_i \equiv \tilde{K}^j{}_{i \mid j}-\frac{2}{3} K_{\mid i}- {\rm M_{Pl}^{-2}} {\cal { J }}_i = 0, 
    \label{eq:CoarseGrainedADM_Con2}
\end{eqnarray}
which \emph{do not} receive any stochastic terms on the r.h.s. Indeed, inserting the coarse grained variables in the linearized l.h.s leads to complete cancellations and no source terms appear. This is an important property of the stochastic augmentation of the ADM system of equations: the dynamics can be seen as a continuous succession of ``drift and kick" evolution steps or a {``succession of small time-step initial condition problems"} \cite{salopek_stochastic_1991}. The l.h.s. of the ADM equations \eqref{eq:CoarseGrainedADM_Dyn3} - \eqref{eq:CoarseGrainedADM_Dyn1} - \eqref{eq:CoarseGrainedADM_Dyn2} preserves the constraints if initial conditions satisfy them. Indeed, satisfying the constraints is an important consideration in any scheme for setting initial conditions in Numerical Relativity. Since the stochastic source terms can be interpreted as  modifying the ``initial conditions'' for each successive time step, their addition must not alter the ADM constraints. Our approach for defining these stochastic source terms satisfies the constraints at first order \new{in perturbation theory} by construction.

The stochastic equations derived here rely on linearized perturbations for determining the noise terms and this is an approximation that underlies practically all derivations of stochastic equations for inflation. However, \new{once a stochastic kick has been given}, the subsequent evolution of the long wavelength configuration is fully non-linear. Overall, this ``linearized noise" approximation together with neglecting the quantum contributions to the interactions (expected to hold from horizon crossing \cite{Launay2024bis}) and the stochastic backreaction of IR fields on the UV modes that determine the noise via \eqref{eq:ReqISTORIZ1}, are the only approximations left in the above equations. 

Equations \eqref{eq:CoarseGrainedADM_Dyn3} - \eqref{eq:CoarseGrainedADM_Dyn1} - \eqref{eq:CoarseGrainedADM_Dyn2} and \eqref{eq:CoarseGrainedADM_Con1} - \eqref{eq:CoarseGrainedADM_Con2}, solved simultaneously with \eqref{eq:ReqISTORIZ1} which defines the noise sources, constitute our dynamical equations for Stochastic Inflation in Numerical Relativity. 

\setlength{\tabcolsep}{4pt}   
\renewcommand{\arraystretch}{1.7}
\begin{table}[]
    \centering
    \resizebox{\linewidth}{!}{
    \begin{tabular}{|c||l|l|}
        \hline
         Potential& Quadratic & Higgs inflection \cite{Hamada14,Bezrukov14} \\
         \hline
         Expression 
         & $V[\phi] = \frac{1}{2}m^2\phi^2$
         & $V[\phi] = \frac{\Lambda v^4\phi^2(3\phi^2 + 2\sqrt{2}\phi v + 6v^2)}{3(3\phi^2 + 2v^2)^2}$ \\
         \hline
         Parameters & $\varepsilon_1 = 0.01$ & $\Lambda = 1.86\times 10^{-6}$ \\
          & $\varepsilon_2 = -0.02$ &  $v = 0.198{\rm M_{Pl}}$\\
         & $H_0 = 10^{-5}{\rm M_{Pl}}$ &  $\phi_0 = -1.2{\rm M_{Pl}}$ \\
         & $m=m(\varepsilon_1,\varepsilon_2,H_0)$
         & $\varepsilon_1=\varepsilon_1(\phi_0)$  \\
         & $\phi_0=\phi_0(\varepsilon_1,\varepsilon_2,H_0)$ 
         & $H_0 = H_0(\varepsilon_1, \phi_0)$\\
         \hline
         Simulation & ${\cal N}^{\circledast}-{\cal N}_0 \sim 5$ & ${\cal N}^{\circledast}-{\cal N}_0 \sim 3$ \\
         & ${\cal N}_f-{\cal N}^{\circledast} \sim 8$& ${\cal N}_f-{\cal N}^{\circledast} \sim 10$ \\
         \hline
    \end{tabular}
    }
    \caption{Potentials used in this study and their parameters, including initial conditions, determined either by \textit{potential-position} or $H_0$ fixing \cite{Launay2025}. ${\cal N}_0$, ${\cal N}^{\circledast}$, and ${\cal N}_f$ are the Bunch-Davies initialisation time, the start of the NR simulation, and the final time in efolds.}
    \label{tab:scenarios}
\end{table}

\begin{figure}[!ht]
    \centering
    \begin{subfigure}{0.23\textwidth}
\includegraphics[width=\linewidth]{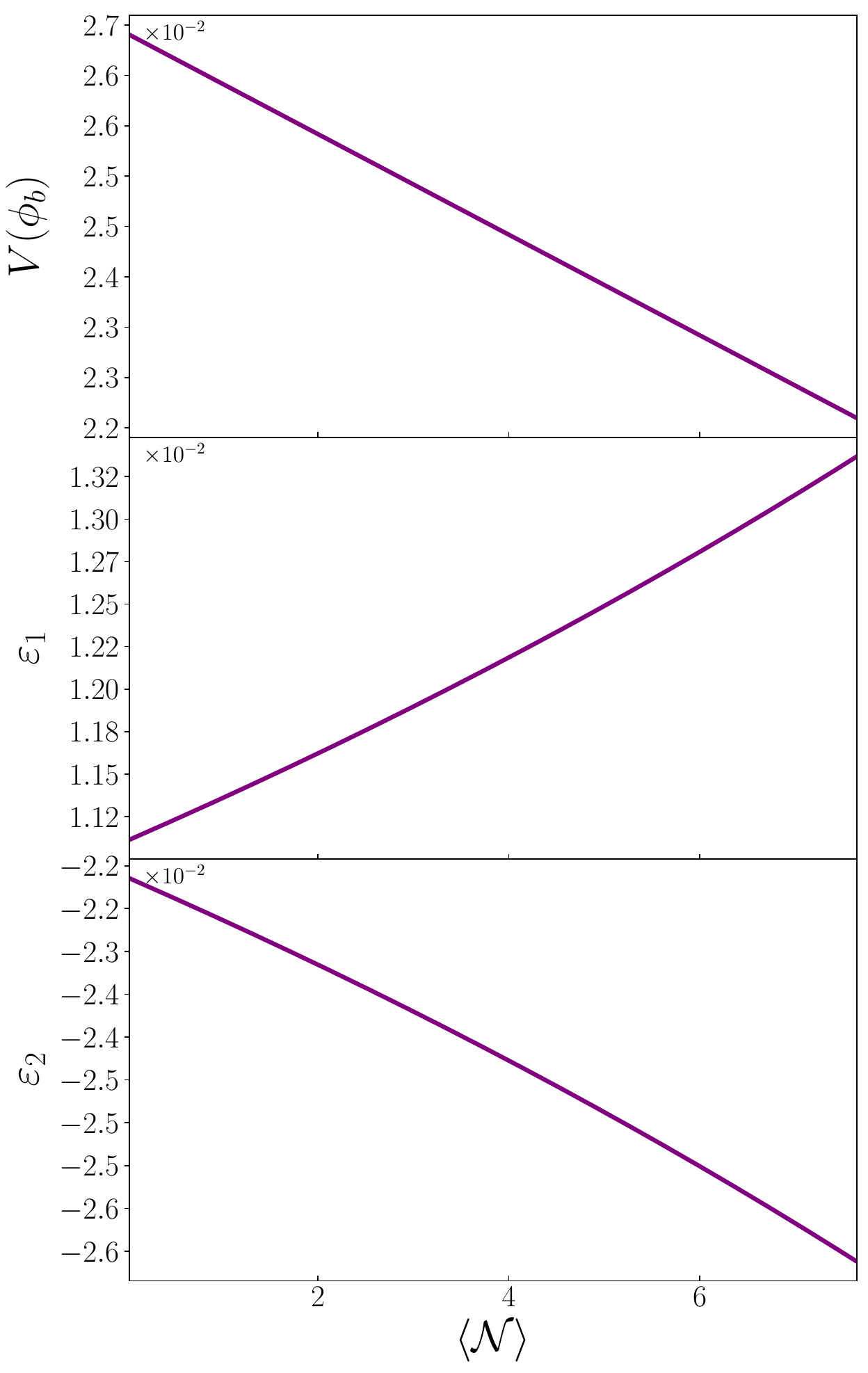}
\caption{Quadratic inflation}
\label{fig:MeanRollsquad}
    \end{subfigure}
    \begin{subfigure}{0.23\textwidth}
\includegraphics[width=0.987\linewidth]{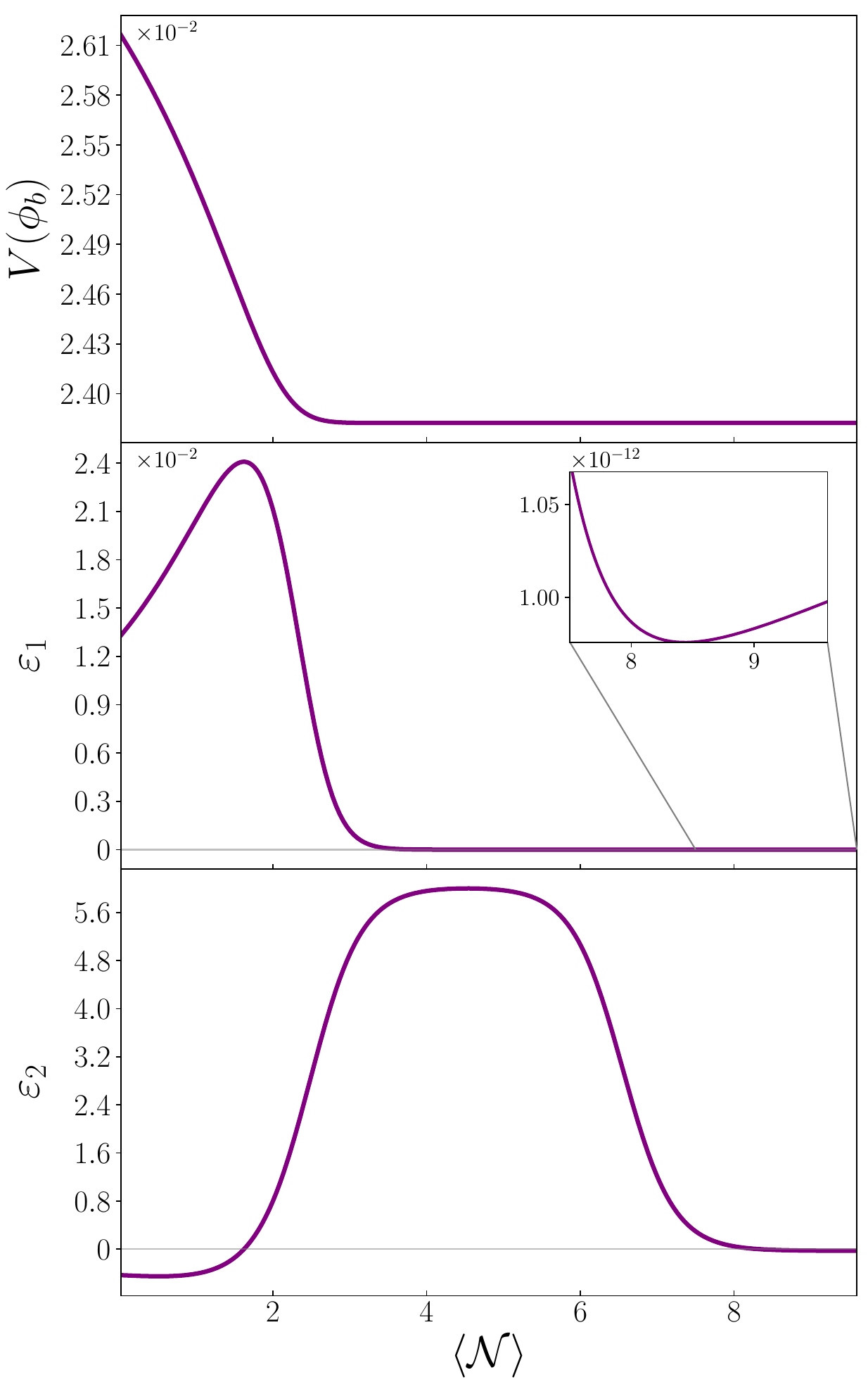}
\caption{Inflection inflation}
\label{fig:MeanRollsusr}
    \end{subfigure}
\caption{\justifying Time background evolution of the potential value (top), first (middle) and second (bottom) slow-roll parameters for both our validation scenarios. Perfect agreement between Friedmann and NR evolutions. ${\rm M_{Pl}}=100$.}
\label{fig:MeanRolls}
\end{figure}

\begin{figure}[!htbp]
    \centering
\includegraphics[width=0.8\linewidth]{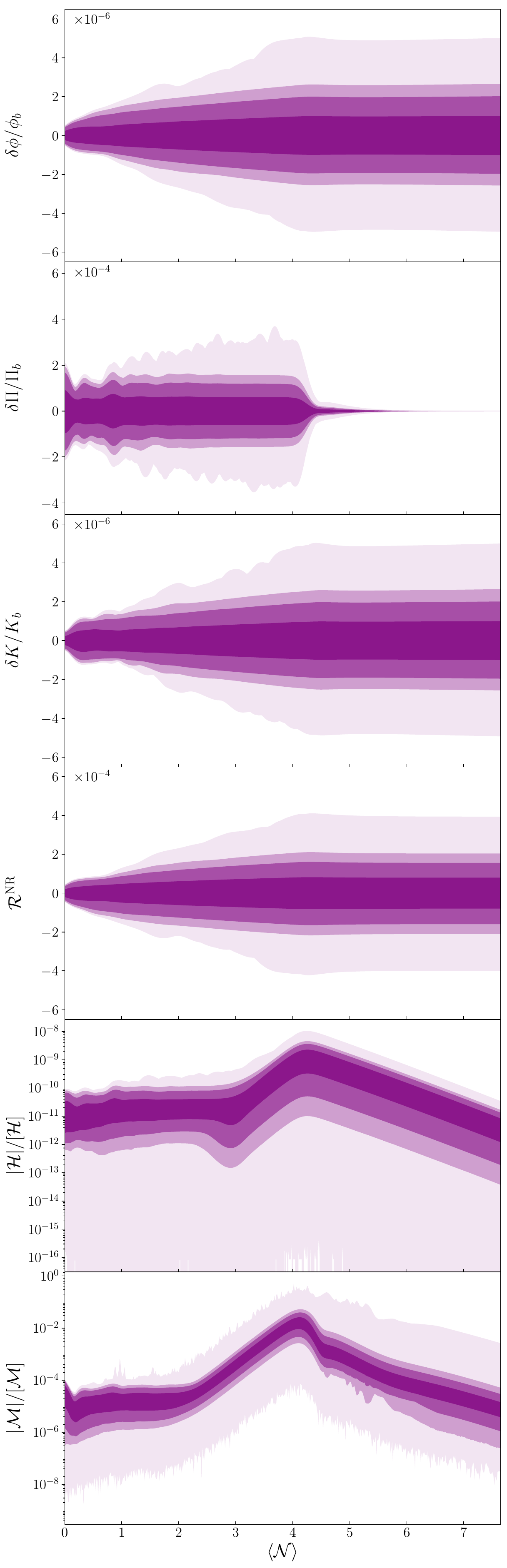}
    \caption{\justifying Contrast of the inflaton (top), its momentum, the extrinsic curvature, the reconstructed ${\cal R}^{\rm NR}$ and the relative Hamiltonian and Momentum constraints (bottom) extracted from the NR simulation of quadratic inflation using $68$, $95$, $99$ and $100$ percentile contours (dense to light purple). ${\rm M_{Pl}}=100$.}
    \label{fig:quad_contrast}
\end{figure}

\begin{figure*}[!t]
    \centering
\includegraphics[width=0.9\linewidth]{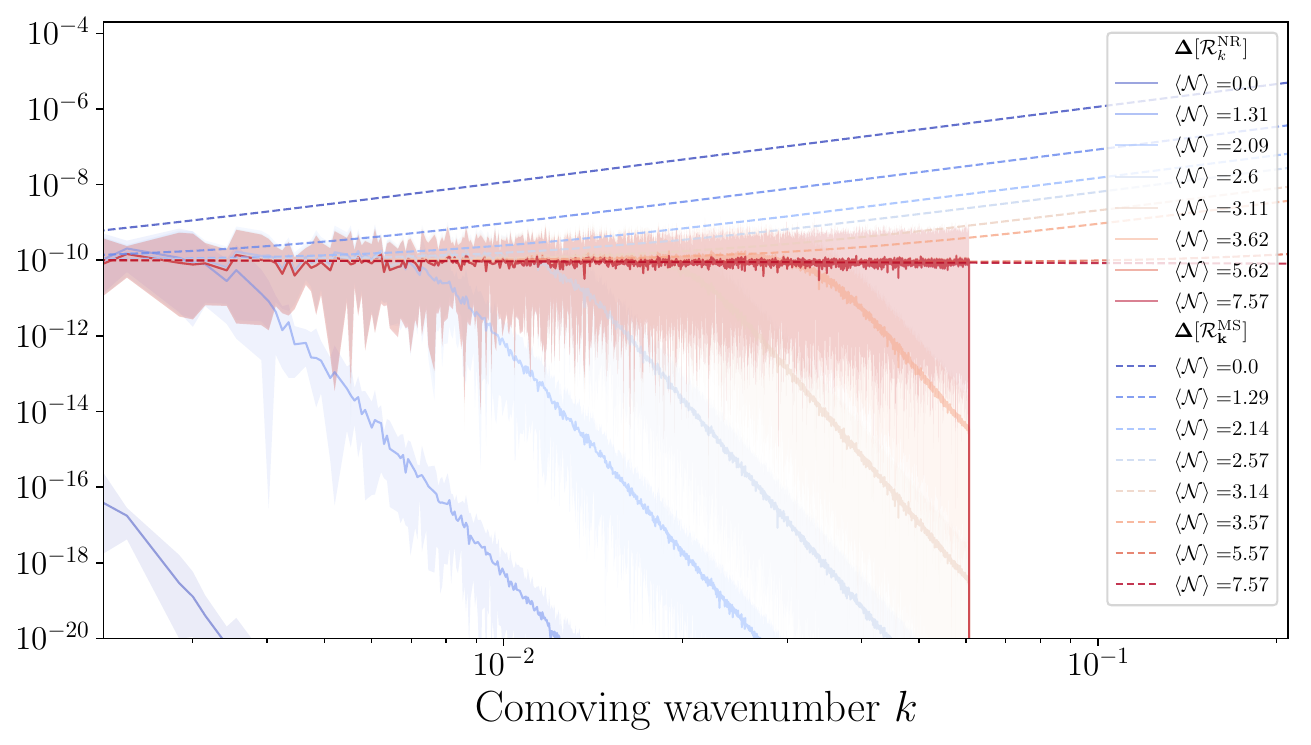}
    \caption{\justifying Linear (dashed, ${\cal R}^{\rm MS}$) and reconstructed (solid, ${\cal R}^{\rm NR}$) binned dimensionless spectra ($\Delta[\cdot]= \frac{k^3}{2\pi^2} |\cdot|^2$) in the quadratic inflation case from initial (navy blue) to final (red) times. Contours account for the full range of points in each bin. Fourier bins are those of the lattice with more than 10 realisations. ${\rm M_{Pl}}=100$.}
    \label{fig:quad_spec}
\end{figure*}

\begin{figure}[!htbp]
    \centering
\includegraphics[width=0.8\linewidth]{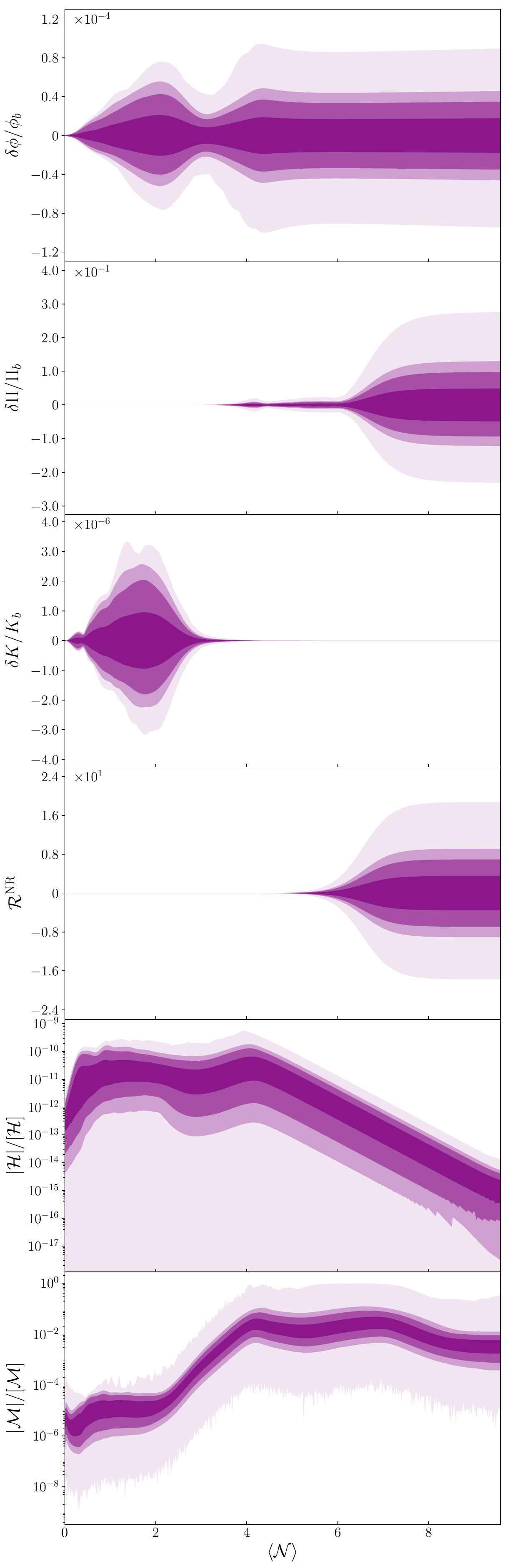}
    \caption{\justifying Contrast of the inflaton (top), its momentum, the extrinsic curvature, the reconstructed ${\cal R}^{\rm NR}$ and the relative Hamiltonian and Momentum constraints (bottom) extracted from the NR simulation of inflection inflation using $68$, $95$, $99$ and $100$ percentile contours (dense to light purple). ${\rm M_{Pl}}=100$.}
    \label{fig:usr_contrast}
\end{figure}

\begin{figure*}[!t]
    \centering
\includegraphics[width=0.9\linewidth]{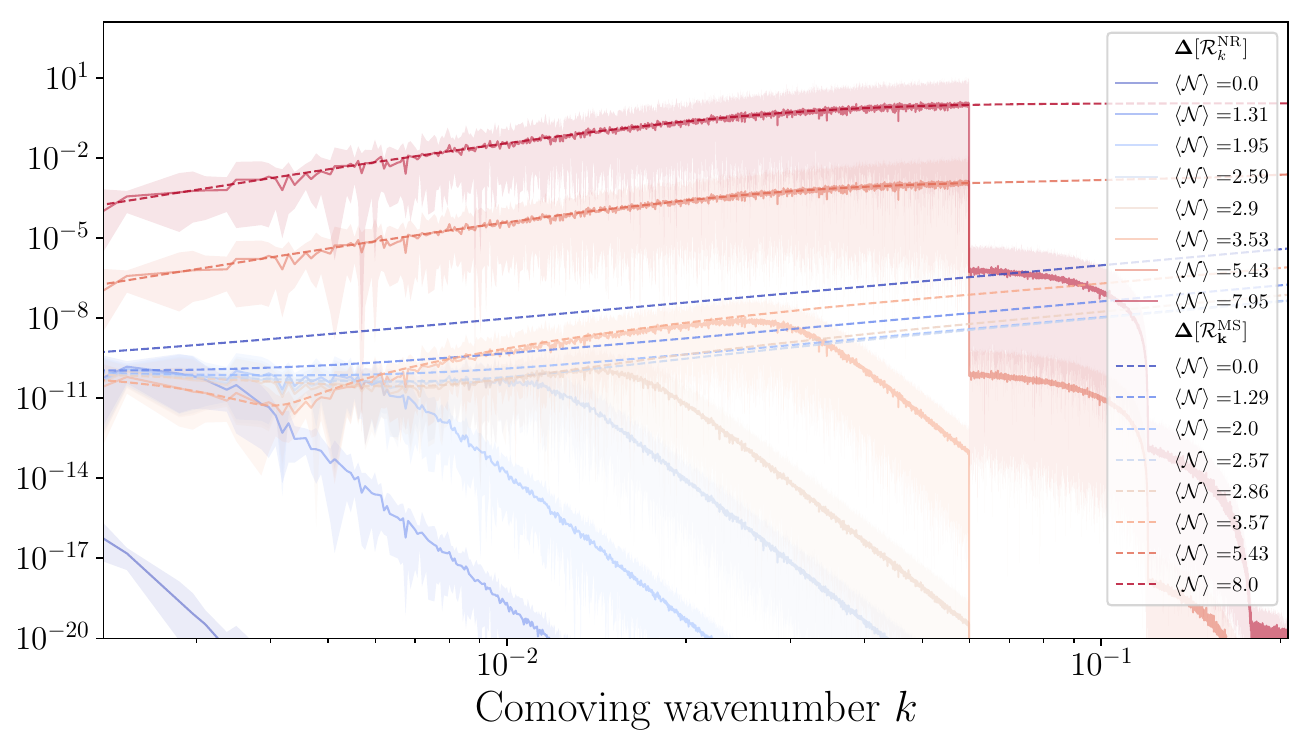}
    \caption{\justifying Linear (dashed, ${\cal R}^{\rm MS}$) and reconstructed (solid, ${\cal R}^{\rm NR}$) binned dimensionless spectra ($\Delta[\cdot]= \frac{k^3}{2\pi^2} |\cdot|^2$) in the inflection inflation case from initial (navy blue) to final (red) times. Contours account for the full range of points in each bin. Fourier bins are those of the lattice with more than 10 realisations. ${\rm M_{Pl}}=100$.}
    \label{fig:usr_spec}
\end{figure*}


\section{Stochastic Numerical Relativity \label{sec:istoriz}}

\subsection{Numerical Pipeline}
Numerically solving eqns.~\eqref{eq:CoarseGrainedADM_Dyn3}, \eqref{eq:CoarseGrainedADM_Dyn1} and \eqref{eq:CoarseGrainedADM_Dyn2} from initial data, while satisfying \eqref{eq:CoarseGrainedADM_Con1} and \eqref{eq:CoarseGrainedADM_Con2}, requires a well-posed reformulation leading to the BSSN equations and variables \cite{BSSN_BS,BSSN_SN} recalled in Appendix \ref{app:BSSN}. These equations along with \eqref{eq:ReqISTORIZ1} for the noise amplitude are the main equations of our stochastic inflation formulation for numerical relativity (NR). To our knowledge, {this is the first code of this kind.}

Before incorporating any stochasticity, multiple codes are available to solve the BSSN system. In this work, we chose to use the most recent \textsc{GRTeclyn} \cite{GRTeclyn}\footnote{Note that an official public release is not yet available, but the software is open-access and reproduces the scalar sector of \cite{Launay2025} against the well established \textsc{GRChombo} \cite{Clough_2015,Andrade2021}.}. It inherits the \textsc{AMReX} software \cite{amrex} within its core, which enables the efficient scalable use of GPU offloading and adaptive mesh refinement (AMR) to solve generic PDEs with finite volume methods. In this work, the numerical scheme of the time evolution is a $4$th order Runge-Kutta (RK$4$).

With \textsc{GRTeclyn} as its core, \textsc{ISTORIz}\footnote{ \textsc{I}nflationary \textsc{STO}chastic \textsc{R}elativity \textsc{I}ntegrator for \textsc{z}eta.
} enables the specific study of inflation together with the stochastic sources derived above. In its current version, \textsc{ISTORIz} sets NR parameters to fairly simple values. The AMR is switched off together with the damping of the constraints (CCZ4 \cite{CCZ41,CCZ42}), the gauge choice is that of geodesic slicing (cosmic time, $\alpha=1$, $\beta^i=0$), and we use periodic boundary conditions. Up to caveats of studying cosmological spacetimes in NR \cite{aurrekoetxea24}, these only constitute arbitrary choices to which the software is not limited.  Given the spectral nature of our stochasticity, using \textsc{AMReX}-based software, such as \textsc{GRTeclyn}, has additional advantages because of the built-in off-loaded Fast Fourier Transform (FFT) capabilities. Key components of \textsc{ISTORIz} are:
\begin{itemize}
    \item[i)] Reading of inflationary background initial conditions ($\phi_b^\circledast$, $\Pi_b^\circledast$, $K_b^\circledast$, $X_b^\circledast$) and spectral initial conditions for the curvature perturbation on comoving hypersurface (${\cal R}_k^\circledast$ and $\dot{\cal R}_k^\circledast$ for each $k$ of the discrete Fourier lattice). In this work, these are all provided by the \textsc{STOIIC} generator presented in \cite{Launay2025}, using \textit{oscode} \cite{Agocs20} to solve the Mukhanov-Sasaki equation from deep subHubble initialisation (time ${\cal N}_0$) to the start of the simulation (time ${\cal N}^{\circledast}$) when the first mode enters the \new{($\sigma$-dependent) IR sector}.
    \item[ii)]  Random draw of the real space Gaussian discrete fields $\boldsymbol{{\cal R}}^\circledast(\boldsymbol{x})$ and $\partial_t\boldsymbol{{\cal R}}^\circledast(\boldsymbol{x})$ from the input spectra (see methodology in \cite{Launay2025}), and, in parallel to the BSSN evolution, evolution of the Mukhanov-Sasaki eq. \eqref{eq:Req} in real space in its well-posed formulation eq.~\eqref{eq:ReqISTORIZ1}.
     Stochastic backreaction can be switched on in this evolution, by promoting background quantities such as $\phi_b$, $K_b$ or $X_b$ to their local spatial value. This defines one of three currently implemented levels of backreaction \new{(\textit{level 2 backreaction}), as discussed and compared in Appendix \ref{app:stochback}.}
    \new{\item[iii)] Initialization of the coarse-grained ($\phi^{\circledast,>}$, $\Pi^{\circledast,>}$, $K^{\circledast,>}$, $X^{\circledast,>}$) at the initial simulation time $t^{\circledast}$ by applying the procedure used in \cite{Launay2025} to the coarse-grained initial conditions $\big(W*\boldsymbol{{\cal R}}\big)(t^\circledast,\boldsymbol{x})$ and $\partial_t\big(W*\boldsymbol{{\cal R}}\big)(t^\circledast,\boldsymbol{x})$. \new{This is necessary since there are minor but non-zero contributions from $W$ being close to, but not exactly $0$ at $t^{\circledast}$ for the modes of interest. If this is not accounted for, we observe a detectable lack of accuracy in the spectrum of the first (low $k$) modes entering the simulations.}}
    \new{\item[iv)] At each RK4 time step, the discrete stochasticity is given by
        \begin{itemize}
            \item[a)] Fourier transform of $\boldsymbol{{\cal R}}(t,\boldsymbol{x})$ and $\partial_t\boldsymbol{{\cal R}}(t,\boldsymbol{x})$
            \item[b)] Integration $\int_t^{t+\Delta t}\boldsymbol{{\cal S}}_{\left\{\Pi, K, A_{ij}\right\}}(u,k)du$
            \item[c)]Inverse Fourier Transform  ${\cal F}^{-1}\{\cdot\}$
            \item[d)] Addition of the stochastic noise to each r.h.s. of the discretised BSSN. 
        \end{itemize}
        This ensures the satisfaction of the constraints up to ${\cal O}(\text{noise}^2)$.
        The stochastic scheme is detailed in Appendix \ref{app:schemes}.}
\end{itemize}

\subsection{Design of the discrete stochastic spacetime \label{sec:designlattice}}
The choice of a geodesic gauge implies that our lattice is a comoving space one. In these coordinates, the Hubble radius $R_H = (aH)^{-1}$ shrinks so that big scales receive stochastic fluctuations first, followed by smaller ones until the lattice's smallest visible scale ($dx$) is reached. In this work, each mode $k$ evolved by eq. \eqref{eq:Req} or \eqref{eq:ReqISTORIZ1} enters the GR evolution via the stochastic terms when $k_\sigma(t)\equiv\sigma R_H^{-1} (t)$ has increased up to $k$. This is practically the case when the window function is set to the rolling rectangle $\Theta[k_{\sigma}(t)-k]$. However, to demonstrate the generality of our framework and resolve all involved time derivatives of $W$, we pick a smoother window function with activation centered on the crossing \begin{equation}\label{eq:ourWindow}
W_k(t)=\frac{1}{2}\Bigg(1+\tanh\Big[s\log\Big(\frac{k_\sigma(t)}{k}\Big)\Big]\Bigg),
\end{equation}
where $s=5.0$ is an arbitrary sharpness parameter\footnote{This should not have an impact on the result in the late-time limit, as long as a window follows requirements such as those defined in \cite{Winitzki00}. However, for a given discrete scheme, not all windows are suitable. In particular the sharpness should be appropriate for the time stepping and for keeping all terms within perturbation theory.}.
Thanks to a random draw performed in Fourier space and the Fourier transforms at every time step, we effectively input non-Markovian noise when such a non-sharp window function is used. 

Let $N$ now be the number of cells per spatial dimension of the simulation's cubic patch of universe and $L = Ndx$ the unitfull comoving size of the latter. Assuming real fields and the usual isotropy of the fluctuations, the range of Fourier modes which can be added is thus made of all norms $k$ such that
$ k=dk\sqrt{a^2+b^2+c^2},$ where $(a,b,c) \in \llbracket 0,\lfloor N/2\rfloor \rrbracket^3$ and $dk \equiv k_{\text{min}} = 2\pi L^{-1}$. In this setting, \textsc{STOIIC} picks $L$ so that the first mode enters when the simulation starts, which is
$L k_{\sigma}^\circledast = {2\pi}$.
\new{For a good approximation of the gradient stencils ($6$th order in this work), our UV cutoff for the fluctuations is taken lower than the Nyquist frequency by imposing $k^{\cal R}_{\text{max}} =  N k_{\sigma}^\circledast/4$ as the last and highest-$k$ mode entering the GR simulation.\footnote{Note that Kreiss-Oliger dissipation is also available in \textsc{GRTeclyn} to damp out high frequency modes, but not deployed here.}}

Similarly, the design of the time dimension in the lattice or stepping is intrinsically linked to the amount of physical time we are interested in and to the available computational resources constraining the choice of $N$. This is explained in Appendix~\ref{app:theta}.

\subsection{Validation}
In the following, we work with the scaling $a^\circledast=1$, units such that ${\rm M_{Pl}} = 100$ and lattices of size $N=256$. The coarse-graining is fixed to $\sigma=1$ for all scenarios. 
Similarly to our previous work \cite{Launay2025}, we study both a vanilla example of inflation for the sake of diagnostics, and a more challenging example of an ultra-slow roll model. Reserving a more thorough investigation for future work, including different types of inflationary histories, we simply report here on the validation and potential of our stochastic method. The two scenarios studied are presented in Table \ref{tab:scenarios} together with their extracted background evolution for the relevant e-fold span in Figure \ref{fig:MeanRolls}. {Note that stochastic backreaction is switched on but its ablation yields minor spectral changes in those two scenarios: we find a detectable but $\lesssim  1$\textpertenthousand\text{} relative change in power spectrum for all scales, see Appendix \ref{app:stochback}.}

Figures~\ref{fig:quad_contrast} and \ref{fig:quad_spec} report the evolution of a simple quadratic inflation scenario and thus slow-roll (SR) dynamics. Modes enter the simulation for about $3.5-4$ efolds as can be seen for the growth of the contrasts in Figure~\ref{fig:quad_contrast}. Afterwards, quantities such as ${\cal R}$ freeze as expected in the super-Hubble regime. The dynamics of such vanilla scenario are well-known and a good test that linear physics is recovered. This is particularly visible in the comparison of the spectrum extracted from the NR simulation ${\cal R}^{\rm NR}$ and the Mukhanov-Sasaki solution ${\cal R}^{\rm MS}$ in Figure~\ref{fig:quad_spec}, once the window has fully rolled from the IR to the UV. Note that the extraction of ${\cal R}^{\rm NR}$ from the evolved BSSN variables was detailed in Appendix B of \cite{Launay2025} and involves i) substraction of the box background, ii) reconstruction of $\Phi$ from BSSN conformal factor and metric, and iii) the linear combination 
\begin{equation}
{\cal R}^{\rm NR} = \Phi -\frac{\langle K \rangle}{3\langle \Pi \rangle}(\phi-\langle \phi\rangle)
\end{equation}
in this gauge. 

An ultra slow-roll phase (USR), $\varepsilon_1  \sim 0$ and $\varepsilon_2 \sim 6$ in our notations, is famously known for enhancing the power spectrum of modes after crossing the horizon \cite{Leach01}. In general, slow-roll-breaking regimes can be reached using features of various forms in the inflaton potential, see, for instance, \cite{Mishra19}. Our potential is motivated by Higgs inflation \cite{Bezrukov14,Hamada14} and has been slightly tuned to show interesting physics within a few efolds.

In this work, the lattice contains modes entering the IR system during the USR phase as well as during the transition from SR to USR, a numerically challenging regime because of the extreme slow-down of the field and the importance of the gradients. This has been at the heart of issues and attempts to fix them in gradient-agnostic approaches using the separate universe approximatio and thus for $\delta N$ formalisms \cite{Artigas24} or standard stochastic inflation \cite{jackson2023,Briaud25}. Our simulations are run with $\sigma = 1$ and are fully gradient-enabled, which is unprecedented in a stochastic framework, whether it is for USR or not.

One-point distributions for real space inhomogeneities in this USR scenario are displayed in Figure~\ref{fig:usr_contrast}, where we report the contrasts of key quantities. The top three panels show the progressive entry of all modes with their full amplitude within about $4$ efolds. Note the double feature structure of the $\phi$ contrast evolution in juxtaposition with that of the extrinsic curvature $K$, which lacks the second growth phase. Qualitatively, field fluctuations that were added in the slow roll phase ($\mathcal{N}\lesssim2$), decay as the field slows down during $2\lesssim\mathcal{N}\lesssim 3$, while at the same time more modes are added as they exit the horizon. These latter modes contribute to the growth seen for $3\lesssim\mathcal{N}\lesssim 4$ after which point all available modes in the simulation have been used. After $\mathcal{N}\sim 3$ the almost constant part of the potential has been reached and $\phi$ diffuses on it without affecting the expansion rate. $K$ therefore does not receive any further perturbations since $\phi$ perturbations no longer change the energy density, leading to the lack of a second growth feature in $K$. This picture is consistent with \cite{prokopec_ensuremathdeltan_2021, Prokopec:2025uvz}    

From $\mathcal{N} \sim 3.5$ USR starts and most of the modes making up $\mathcal{R}^{\rm NR}$ are already super-Hubble and undergo an exponential enhancement (expected for USR) until $\mathcal{N}\sim 8$. After that point, a slow roll phase commences with $\left(\varepsilon_1,\varepsilon_2\right)\simeq \left(0,-2.8 \times  10^{-2}\right)$ and  $\mathcal{R}^{\rm NR}$ freezes as expected. This phase occurs after a very shallow maximum is passed and the field starts rolling down the potential with $dV/d\phi<0$.  Note that $\mathcal{R}^{\rm NR}$ reaches the non-perturbative values of $\sim 10$ for the furthest outliers in real space. The contrasts $\delta\Pi/\Pi_b$ also become relatively large at late times in sharp contrast to the quadratic case. 

Up to a certain maximum inhomogeneity amplitude (usually defined for dimensionless power spectrum as $\Delta[{\cal R}]\sim 1$ \cite{Franciolini24}), and due to the $\varepsilon_1\to 0$ limit, USR is usually assumed to be well-described by the linear solution from the Mukhanov-Sasaki equation, despite reaching non-perturbative amplitudes. This is visible for the spectra displayed in Figure~\ref{fig:usr_spec}, which, at complete overlap of the closed linear and open IR non-linear system, shows excellent agreement.\footnote{This can also be confirmed by extracting the non-linear generalisation of ${\cal R}$, introduced in \cite{Launay2025}, showing no difference.} {We find that, despite reaching up to $\Delta[{\cal R}]\sim 0.1-1$, departure from the final linear spectral prediction is seemingly not as strong as} \cite{caravano2024usr}. In the future, a fully quantitative study will make detailed comparisons in this limit.

We note the last two panels of Figures \ref{fig:quad_contrast} and \ref{fig:usr_contrast} as a crucial verification of both our numerical and theoretical pipeline. In fact, we claimed satisfaction of the constraints at first order in the noise. To evaluate the order of the terms in the constraints, we define  the \textit{absolute constraint magnitudes} $[{\cal H}]$ and $[{\cal M}]$, calculated by summing the absolute value of the individual terms in the constraints \new{(see eq. \ref{eq:abscons} in Appendix \ref{app:conv} for a reminder of their definition)}.
In its entry phase, the noise in ${\cal R}$ has a maximum amplitude of $\sim 10^{-5}$. In both scenarios, the Hamiltonian constraint shows violation at the next order in perturbation theory ($\sim 10^{-10}$) when compared to the absolute magnitude $[{\cal H}]$. After that, inflation naturally damps the constraint. 

For the Momentum constraint, being null at $0th$ order, the absolute magnitude $[{\cal M}]$ is the first order and we thus observe a second order violation as well in this phase. As a consistency check and similarly to \cite{Launay2025}, we find that switching any sign in our noise contributions yields a violation at first order (meaning ${\cal H}/[{\cal H}]\sim 10^{-5}$ and  ${\cal M}/[{\cal M}]\sim 1$). In the quadratic case, the last mode crossing makes the $|{\cal M}|/[{\cal M}]$ momentarily a bit higher than second order because of the gradient resolution, which is acceptable given that a smaller $k_{\text{max}}$ would fix that. A similar conclusion can be reached for the inflection case where $|{\cal M}|/[{\cal M}]$ sits later around $10^{-2}$, which is acceptable given inhomogeneity of order $10$ in ${\cal R}$ at that point. 

\new{Finally, a positive convergence analysis for the constraints and the spectra is reported in Appendix \ref{app:conv}.} This completes the validation and physicality check of our method, making it ready for future work on both fully non-perturbative and non-linear experiments for which no analytical results are known.


\subsection{A first look at anisotropic DOFs \label{app:Aij}}
\begin{figure*}[!ht]
    \centering
    \begin{subfigure}{\textwidth}
\includegraphics[width=\linewidth]{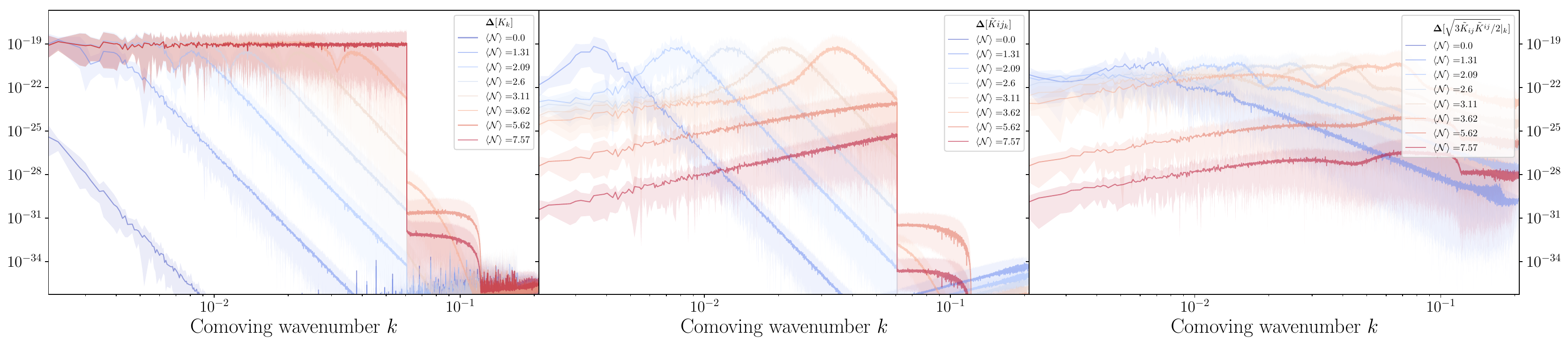}
\caption{Quadratic inflation}
    \end{subfigure}
    \begin{subfigure}{\textwidth}
\includegraphics[width=\linewidth]{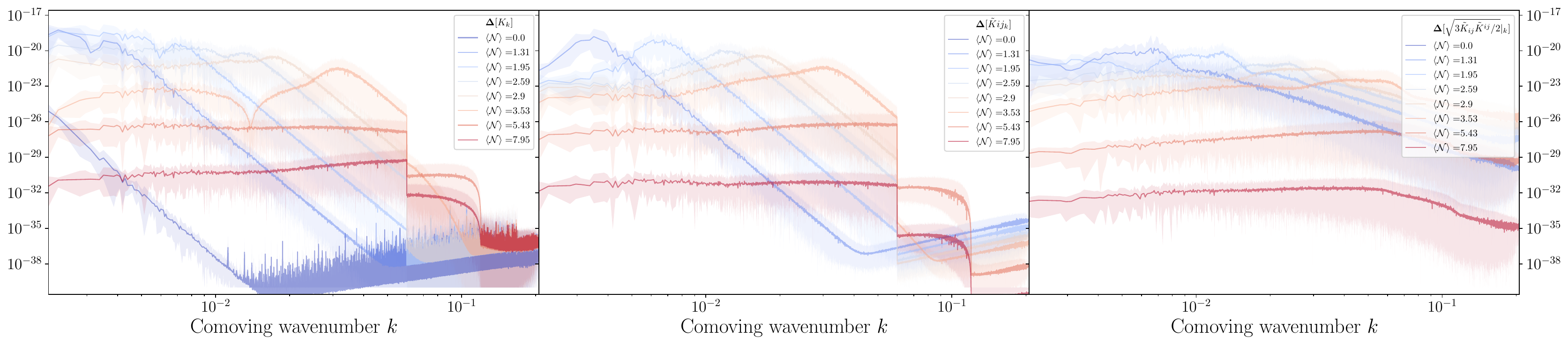}
\caption{Inflection inflation}
    \end{subfigure}
\caption{\justifying Binned dimensionless spectra ($\Delta[\cdot]= \frac{k^3}{2\pi^2} |\cdot_k|^2$) of $K$, $\tilde{K}$, and $\sqrt{3\tilde{K}_{ij}\tilde{K}^{ij}/2}$ in the quadratic (top) and inflection inflation case (bottom) from initial (navy blue) to final (red) times. Contours account for the full range of points in each bin. Fourier bins are those of the lattice with more than 10 realisations. ${\rm M_{Pl}}=100$.}
\label{fig:specAK}
\end{figure*}
As a first demonstration of the potential of such simulations beyond standard stochastic inflation, we analyse the evolution of the anisotropic degrees of freedom encoded in $\tilde{K}_{ij}$. In future work, applying the transverse projector will yield the first extraction of scalar-induced gravitational wave power from a Bunch-Davies vacuum evolved in stochastic NR. We reserve these investigations of PGWs for future work and focus on the general anisotropic amplitude in our two scenarios.

Discarding the anisotropic degrees of freedom of General Relativity is central to the separate universe approximation (SUA) \cite{wands_new_2000} and the Hamilton-Jacobi approach \cite{salopek_nonlinear_1990, rigopoulos_separate_2003, prokopec_ensuremathdeltan_2021} { - but see \cite{Tanaka:2023gul,Tanaka:2024mzw}}. With a fully relativistic evolution of Bunch-Davies fluctuations, we can confirm if this is appropriate and understand the validity of such approximations more quantitatively. We know of course that $K$'s background ($k=0$ mode) is dominating overall given that $\tilde{K}$ has no background order contribution. We are interested in the perturbations at first and second order in perturbation theory, in particular in the case of USR where ${\cal R}$ blows up. For that purpose Figure \ref{fig:specAK} shows the spectra of $K$, $\tilde{K}$ (taken as $\tilde{K}_{ij}(k)\tilde{K}^{ij}(k)^*$ in Fourier space) and $\sqrt{3\tilde{K}_{ij}\tilde{K}^{ij}/2}$. 

It is of interest to compare the anisotropic $(-\tilde{K}_{ij}\tilde{K}^{ij})$ and isotropic ($\frac{2}{3}K^2$) contributions to the Hamiltonian constraint, given that the latter is the only one commonly considered in the SUA. Note that $K$ and $\tilde{K}$ (not squared) do not compete in any of the ADM equations without being squared or contracted respectively, see eqns.~\eqref{eq:CoarseGrainedADM_Dyn3} - \eqref{eq:CoarseGrainedADM_Dyn1} - \eqref{eq:CoarseGrainedADM_Dyn2} 
and
\eqref{eq:CoarseGrainedADM_Con1}, apart from the Momentum constraint \eqref{eq:CoarseGrainedADM_Con2}, usually identified as the culprit hampering such approximation \cite{cruces_stochastic_2022,Artigas25}. 

Looking at the left panels first, we find that perturbations in $K$ freeze in the quadratic slow-roll case but decay in the USR case. This is also visible in the contrasts present in Figures \ref{fig:quad_contrast} and \ref{fig:usr_contrast}. Switching to the second panels, we see that anisotropic and isotropic powers can be comparable in magnitude in both scenarios initially but the former decays quickly after horizon crossing and are thus negligible in the quadratic case. In the USR case, perturbations in $K$ decay as well, implying that all terms in $K_{ij}$ are of similar magnitude and that the SUA truncation of the momentum constraint could potentially be violated. 

The last panel probes most relevant terms competing with $K$: these are expected to be of second order given that $\tilde{K}$ has no background order, and, in fact, a careful observation shows anisotropic spectra to be subdominant to the isotropic modes for most, although not all, times. \new{Clearly, the striking hierarchy between isotropic and anisotropic expansion that characterizes the quadratic case is not present in the USR case.} As pointed out in \cite{Artigas25} through  use of the non-adiabatic pressure, the anisotropic sector cannot be neglected a priori for the satisfaction of the momentum constraint in regimes such as this one.  Further quantitative comparison between the isotropic and anisotropic parts of the extrinsic curvature is left for future work.

\section{Conclusion}\label{sec:Conclusion}

In this work, we pioneered the full evolution of the Bunch-Davies vacuum in inflation using numerical relativity and the stochastic formalism.  In the latter, modes do not all enter the simulation at the same initial time as done in our previous work \cite{Launay2025} (see also \cite{Florio2024} for tensor modes) but at a given elapsed time relative to their Hubble horizon crossing, thus better accounting for quantum diffusion. By achieving this, we provide a framework generalizing stochastic inflation, inflationary numerical relativity and lattice cosmology beyond their standard assumptions.

Two test scenarios were studied, a very perturbative and linear case of slow-roll inflation and a potential with an inflection point triggering ultra slow-roll and so non-perturbative growth. Both scenarios are successfully evolved, matching expectations for the linear spectra and satisfying the Energy and Momentum Constraints for our scales of interest; this is a substantial achievement for a stochastic framework and a strong validation that our pipeline is numerically and theoretically accurate. \new{We were also able to follow the evolution of anisotropic expansion, normally out of reach for stochastic approaches. Indeed, such simulations completely remove the reliance of stochastic inflation on the separate universe approximation and its numerous caveats, the restriction to isotropic expansion, the $\Delta\mathcal{N}$ formalism or the need to pick specific temporal hypersurfaces.}

In the near future we hope to use such simulations to gain more accurate insights on non-linear inflationary physics and models such as ultra slow-roll, and e.g. the related primordial black hole production which have been at the centre of attention, and address debates that have arisen over the past years because of the need for the aforementioned approximations. By including full General Relativity, we are now able to confront previous work as well as giving new signals such as fully non-linear primordial gravitational waves, including scalar-induced ones. For all fields evolved in their semi-classical portion of spacetime, we will also be able to extract non-Gaussianities of a new kind, whether they are additional scalars or from the tensorial sector. Finally, special attention will be given to regimes where quantum-diffusion dominates and where stochasticity is a key component of the dynamics. We therefore hope to provide upcoming experiments with an enlarged testable inflationary landscape.

\begin{acknowledgments}
For their advice, expertise and excellent work on developing core software for the GRTL collaboration, thanks go to J. Kwan and M. Radia, supported by the Intel oneAPI COE and STFC DiRAC. Thanks are also to be given by Y.L. to other colleagues for insightful and supportive conversations, notably E. Florio, C. McCulloch, E. Pajer, S. Renaux-Petel, R. Bond, and E. Tomberg. Y.L. is supported by the STFC DiS-CDT scheme and the Kavli Institute for Cosmology, Cambridge. E.P.S.S. acknowledges funding from STFC Consolidated Grant No. ST/P000673/1. Computational resources were supported by STFC DiRAC HPC Facility (www.dirac.ac.uk) funded by BEIS capital funding via
STFC Capital Grants ST/P002307/1 and ST/R002452/1
and STFC Operations Grant ST/R00689X/1 and a grant from G-research. Y.L. thanks the CoBaLt and Inflation 2025 conference committees and sponsors for their financial support.

\end{acknowledgments}
\bibliography{mybiblio}

\vspace*{\fill}
\appendix 

\section{Gauge invariant perturbations as functionals of ${\cal R}$ \label{app:comProof}}
The scalar linear constraints of general relativity write
\begin{equation}
\left \{
\begin{aligned}
        4\Delta \Phi-4H \kappa-2{\rm M_{Pl}^{-2}}\delta \rho = 0, &\\
         -2\partial_i\left[\Delta \chi+\kappa-\frac{3}{2}{\rm M_{Pl}^{-2}}\frac{\dot{\phi_b}}{\alpha_b}\delta \phi \right]= 0, &
\end{aligned}\right .
    \label{eq:ConstraintsCPT}
\end{equation}
where $\chi \equiv-\frac{a^2}{\alpha_b}(B-\dot{E})$, $\kappa \equiv 3\left(\frac{\dot{\Phi}}{\alpha_b}+H \Psi\right)-\Delta \chi$, and $ \delta\rho = \frac{\dot{\phi_b}}{\alpha_b}\frac{\delta\dot{\phi}}{\alpha_b}+\frac{dV}{d\phi}(\phi_b)\delta\phi-\frac{{\dot{\phi_b}}^2}{\alpha_b^2}\Psi$. 

In the following we set $\alpha_b = 1$, working  in cosmic background time w.l.o.g.
We can chose for convenience the comoving gauge where $\delta\phi^{co}=E^{co}=0$, and one has
\begin{equation}
    \left \{
    \begin{aligned}
        \delta\rho^{co} &= -\dot{\phi_b}^2\Psi^{co},\\
        \chi^{co} & = -a^2B^{co},\\
        \kappa^{co} & = 3(\dot{\Phi}^{co}+H\Psi^{co})+a^2\Delta B^{co},
    \end{aligned}\right .
\end{equation}
together with ${\cal R}^{co} = \Phi^{co}$. Equation \eqref{eq:ConstraintsCPT} thus reduces to
\begin{equation}
    \left\{\begin{array}{cl}
         \Delta {\cal R}^{co} - a^2H\Delta B^{co} & = -\frac{1}{2}{\rm M_{Pl}^{-2}}\dot{\phi_b}^2\Psi^{co}, \\
         \dot{{\cal R}}^{co}+H\Psi^{co} & = 0,
    \end{array}\right .
\end{equation}
which is straightforward to solve in Fourier space
\begin{equation}
    \left\{\begin{array}{cl}
         B^{co} & = a^{-2}H^{-1}{\cal R}^{co}_k+\varepsilon_1k^{-2} \dot{{\cal R}}^{co}_k, \\
         \Psi^{co} & = -H^{-1}\dot{{\cal R}}^{co}_k.
    \end{array}\right .
\end{equation}
Together with $\chi^{co}$ and $\delta\rho^{co}$, all quantities needed to evaluate the linear gauge invariant variables of eq.~\eqref{eq:GIquantities} are known, yielding eq. \eqref{eq:GaugeInvExplicit} in the comoving gauge. These are relations between gauge invariant variables and are therefore also true for any other gauge choice. 

\vspace*{\fill}

\section{BSSN equations \label{app:BSSN}}
The BSSN reformulation of the ADM equations reads
\begin{equation}
\left\{
\begin{aligned}
 \partial_t X -\frac{2}{3} X \alpha K+\frac{2}{3} X \partial_k \beta^k-\beta^k \partial_k X = {\cal F}^{-1}\{\boldsymbol{{\cal S}}_{X}^{BSSN}\},&\\
 \partial_t \tilde{\gamma}_{i j}+2 \alpha A_{i j}-\tilde{\gamma}_{i k} \partial_j \beta^k-\tilde{\gamma}_{j k} \partial_i \beta^k &\\
 \quad+\frac{2}{3} \tilde{\gamma}_{i j} \partial_k \beta^k-\beta^k \partial_k \tilde{\gamma}_{i j} ={\cal F}^{-1}\{\boldsymbol{{\cal S}}_{\tilde{\gamma}_{ij}}^{BSSN}\}, &\\
  \partial_t K+\gamma^{i j} D_i D_j \alpha-\alpha\left(A_{i j} A^{i j}+\frac{1}{3} K^2\right) &\\
-\beta^i \partial_i K-4 \pi \alpha(\rho+S) = {\cal F}^{-1}\{\boldsymbol{{\cal S}}_{K}^{BSSN}\}, &\\
 \partial_t A_{i j}-X\left[-D_i D_j \alpha+\alpha\left(R_{i j}- {\rm M_{Pl}^{-2}}\alpha S_{i j}\right)\right]^{{TF}} &\\
-\alpha\left(K A_{i j}-2 A_{i l} A_j^l\right)
-A_{i k} \partial_j \beta^k-A_{j k} \partial_i \beta^k &\\
+\frac{2}{3} A_{i j} \partial_k \beta^k-\beta^k \partial_k A_{i j} = {\cal F}^{-1}\{\boldsymbol{{\cal S}}_{A_{ij}}^{BSSN}\},&\\
 \partial_t \tilde{\Gamma}^i-2 \alpha\left(\tilde{\Gamma}_{j k}^i A^{j k}-\frac{2}{3} \tilde{\gamma}^{i j} \partial_j K-\frac{3}{2} A^{i j} \frac{\partial_j X}{X}\right) &\\
+2 A^{i j} \partial_j \alpha-\beta^k \partial_k \tilde{\Gamma}^i 
-\tilde{\gamma}^{j k} \partial_j \partial_k \beta^i-\frac{1}{3} \tilde{\gamma}^{i j} \partial_j \partial_k \beta^k &\\
-\frac{2}{3} \tilde{\Gamma}^i \partial_k \beta^k+\tilde{\Gamma}^k \partial_k \beta^i+ 2 {\rm M_{Pl}^{-2}}\alpha \tilde{\gamma}^{i j} S_j = {\cal F}^{-1}\{\boldsymbol{{\cal S}}_{\tilde{\Gamma}}^{BSSN}\}, &\\
 \partial_t \phi-\alpha \Pi-\beta^i \partial_i \phi = 0, &\\
 \partial_t \Pi-\beta^i \partial_i \Pi-\alpha \partial_i \partial^i \phi-\partial_i \phi \partial^i \alpha &\\
-\alpha\left(K \Pi-\gamma^{i j} \Gamma_{i j}^k \partial_k \phi-\frac{d V}{d \phi}\right)= {\cal F}^{-1}\{\boldsymbol{{\cal S}}_{\Pi}^{BSSN}\},  &\\
         \mathcal{H}=R+K^2-K_{i j} K^{i j}- 2 {\rm M_{Pl}^{-2}}\rho = {\cal F}^{-1}\{\boldsymbol{{\cal S}}_{{\cal H}}^{BSSN}\}, & \\
        \mathcal{M}_i=D^j\left(\gamma_{i j} K-K_{i j}\right)- {\rm M_{Pl}^{-2}}S_i ={\cal F}^{-1}\{ \boldsymbol{{\cal S}}_{{\cal M}}^{BSSN}\}, &
    \end{aligned}\right .
        \label{eq:BSSNCoarseGrained}
\end{equation}
where the metric is written as in eq. \eqref{eq:ADMmetric} up to a minus sign convention for the shift $\beta^i$ and where we have introduced the conformal metric and factor, the conformal traceless extrinsic curvature and the contraction of the conformal metric connection
\begin{equation}
\left\{
\begin{aligned}
    \gamma_{i j} & =\frac{1}{X} \tilde{\gamma}_{i j}, \\ X & =\left(\operatorname{det} \gamma_{i j}\right)^{-\frac{1}{3}},\\
    A_{ij} & = X\tilde{K}_{ij},\\
    \tilde{\Gamma}^i& =\tilde{\gamma}^{j k} \tilde{\Gamma}_{j k}^i,
\end{aligned}\right .
    \label{eq:defVarBSSN}
\end{equation}
and where the r.h.s. Fourier transforms were computed in \cite{Launay24} as
\begin{equation}
\left \{
\begin{aligned}
&  \boldsymbol{{\cal S}}_{K}^{BSSN} =-\varepsilon_1\boldsymbol{{\cal S}}_{{\cal R}}.,\\
&  \boldsymbol{{\cal S}}_{A_{ij}}^{BSSN}  =\varepsilon_1(\frac{1}{3}\delta_{ij}-k^{-2}k_ik_j ) \boldsymbol{{\cal S}}_{{\cal R}},\\
& \boldsymbol{{\cal S}}_{\Pi}^{BSSN} = \sqrt{2\varepsilon_1} {\rm M_{Pl}} \boldsymbol{{\cal S}}_{{\cal R}},\\
\end{aligned}\right .
    \label{eq:RHSBSSN}
\end{equation}
and all others being perfectly vanishing.

Note that this work does not implement the tensor sources, derived in \cite{Launay24}.

\section{Numerical and stochastic schemes \label{app:schemes}}
At each time step and for the vector of fields ${\cal X}$, such that $\partial_t {\cal X}= {\cal G}[{\cal X}(t),t,\boldsymbol{x}]= {\rm BSSN}[{\cal X}(t),t,\boldsymbol{x}]+{\cal I}[t,\boldsymbol{x}]$, then the 4th order Runge-Kutta (RK4) scheme writes
$${\cal X}(t+\Delta t, \boldsymbol{x})={\cal X}(t, \boldsymbol{x})+\Bigg(\frac{1}{6}{\cal K}_1+\frac{1}{3}{\cal K}_2+\frac{1}{3}{\cal K}_3+\frac{1}{6}{\cal K}_4\Bigg)\Delta t,$$
where each term writes sequentially as
\begin{equation}
\left \{
    \begin{aligned}
    {\cal K}_1& ={\cal G}[{\cal X}(t),t,\boldsymbol{x}],\\
    {\cal K}_2&={\cal G}[{\cal X}(t)+{\cal K}_1\frac{\Delta t}{2},t+\frac{\Delta t}{2},\boldsymbol{x}],\\
    {\cal K}_3& = {\cal G}[{\cal X}(t)+{\cal K}_2\frac{\Delta t}{2},t+\frac{\Delta t}{2},\boldsymbol{x}],\\
    {\cal K}_4& ={\cal G}[{\cal X}(t)+{\cal K}_3{\Delta t},t+{\Delta t},\boldsymbol{x}].\\
    \end{aligned}\right .
\end{equation}
The subtelty here is the stochastic contribution of the r.h.s. ${\cal I}[t,\boldsymbol{x}] =  {\cal F}^{-1}\{{\cal S}_k(t)\boldsymbol{\alpha_{\vec{k}}}\}$
 writing generically ${\cal S}_k(t) = a^{\cal X}_k(t)\ddot{W}_k(t)+b^{\cal X}_k(t)\dot{W}_k(t)$. Note that with stochastic backreaction, ${\cal I}$ becomes a function of the fields as well. In stochastic PDE language, applying the RK4 method to the stochastic contribution does not give the same convergence for the stochastic part \cite{WarningSRK}, rather that of the Euler–Maruyama method. In the future, we could use a proper stochastic scheme such as those given in \cite{Rossler10} and used in \cite{cruces_stochastic_2022}. However, this would take significant edits in the \textsc{AMReX} core and we expect very little impact if the sources are kept in the perturbative regime compared to the background.

\section{Design of the space to time ratio \label{app:theta}}
Let $\Delta t$ be the time step, $N_T$ the number of time steps, and $\theta = \Delta t/dx$ the Courant factor, which has an upper bound for stability. We are interested in probing the crossing of all modes in our lattice, from $k_{\text{min}}$ to $k^{\cal R}_{\text{max}}$. Let us assume that our simulation's expansion is roughly that of a De Sitter universe with a constant $H^\circledast$, so that $k_{\sigma}(t) \simeq \sigma H^\circledast e^{\cal N}$, where ${\cal N}= H^\circledast(t-t^\circledast)$ is the number of efolds since the start of the simulation. This provides a proxy to compute the number of efolds between the first and the last mode crossing
$${\cal N}^{\otimes}= \ln \bigg (\frac{k_{\text{max}}^{\cal R}}{k_{\text{min}}}\bigg) = \ln \bigg (\frac{N}{n}\bigg).$$
Our simulation needs to last up to that time, together with an amount $\Delta {\cal N}$ of supplementary efolds for the study of the superHubble regime. The feasibility of this study thus depends entirely on our computational resources with respect to the value of $N$.

The choice of the time-to-space ratio $\theta$ is the most constrained one
Beyond the Courant upper bound and stability in general, the choice of $\theta$ is still key. Given the physical duration ${\cal N}_{tot} = {\cal N}^{\otimes}+\Delta{\cal N}$, the number of time steps is
$$N_T \simeq \frac{{\cal N}_{tot}}{H^\circledast \Delta t} = \frac{N\sigma}{2\pi \theta}({\cal N}^{\otimes}+\Delta {\cal N}).$$
In other words, trying to reach an excessively small value for $\theta$ will jeopardize our chances to run the simulation in a reasonable time.

Instinctively, the choice of $\theta$ also controls the number of modes entering the simulation per time step. Indeed, when approximating the shift in the window's range between $t$ and $t+\Delta t$ as $\Delta k_{\sigma}(t) \simeq k_\sigma(t)H \Delta t$, we infer that the ratio to the gap before the next discrete mode is
$$\frac{\Delta k_\sigma(t)}{k_{\text{min}}} \simeq \frac{2\pi\theta}{N\sigma} e^{{\cal N}(t)},$$
which essentially says that the window has an exponentially growing range. For a given ${\cal N}^{\otimes}$ and $\sigma$, one can in particular pick $\theta$ such that no more than one discrete mode enters the lattice per time step, by requiring $\Delta k_\sigma^\otimes\leq k_{\text{min}}$, or equivalently
$$ \theta < \frac{n\sigma}{2\pi},$$
which is more constraining than the stability condition only on superHubble cutoffs ($\sigma <1$).

It is important to note that for $\sigma\geq1$, the simulation will not see more than one discrete mode enter at a time if sticking to the Heaviside windowing or a sharp enough window. Using a smooth window function prevents stochastic spikes from happening in that sense.

\section{\new{Convergence} \label{app:conv}}

\begin{figure}[!ht]
\includegraphics[width=\linewidth]{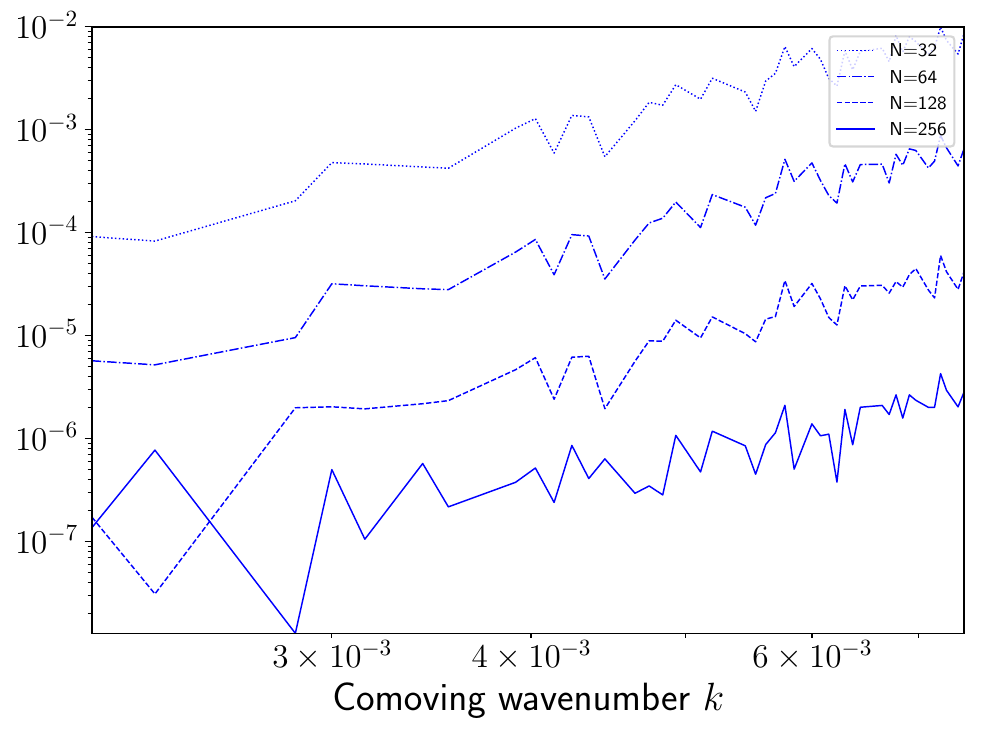}
\caption{\justifying \new{ Convergence test in a linear scenario (quadratic inflation) for the relative difference of the NR curvature perturbation and the Mukhanov-Sasaki equation result $|1-\Delta[{\cal R}^{\rm NR}]/\Delta[{\cal R}]$ across multiple runs with $N=32$ (dotted), $N=64$ (dash-dotted), $N=128$ (dashed), and $N = 256$ (solid). $k_{\text{max}}^{\cal R}/dk$ is set to $8$ across all runs, giving about $\ln 8\simeq 2$ efolds of stochastic sourcing. }}
\label{fig:specConvergence}
\end{figure}

\begin{figure*}[!ht]
    \centering
    \begin{subfigure}{0.45\textwidth}
\includegraphics[width=\linewidth]{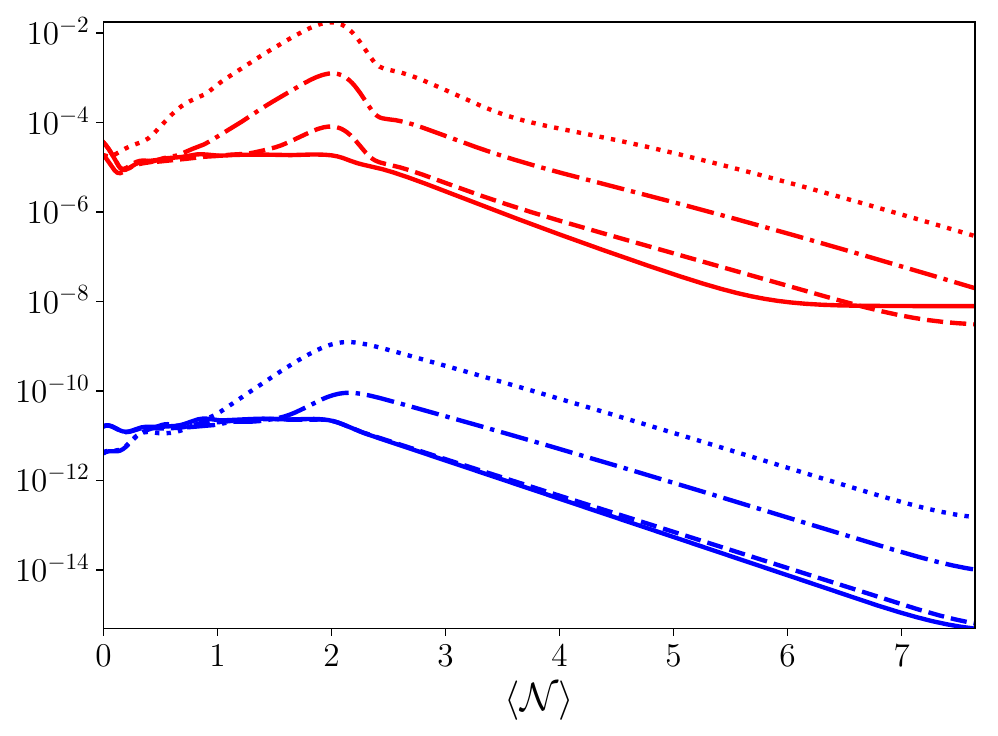}
\caption{Quadratic inflation}
    \end{subfigure}
\begin{subfigure}{0.45\textwidth}
\includegraphics[width=\linewidth]{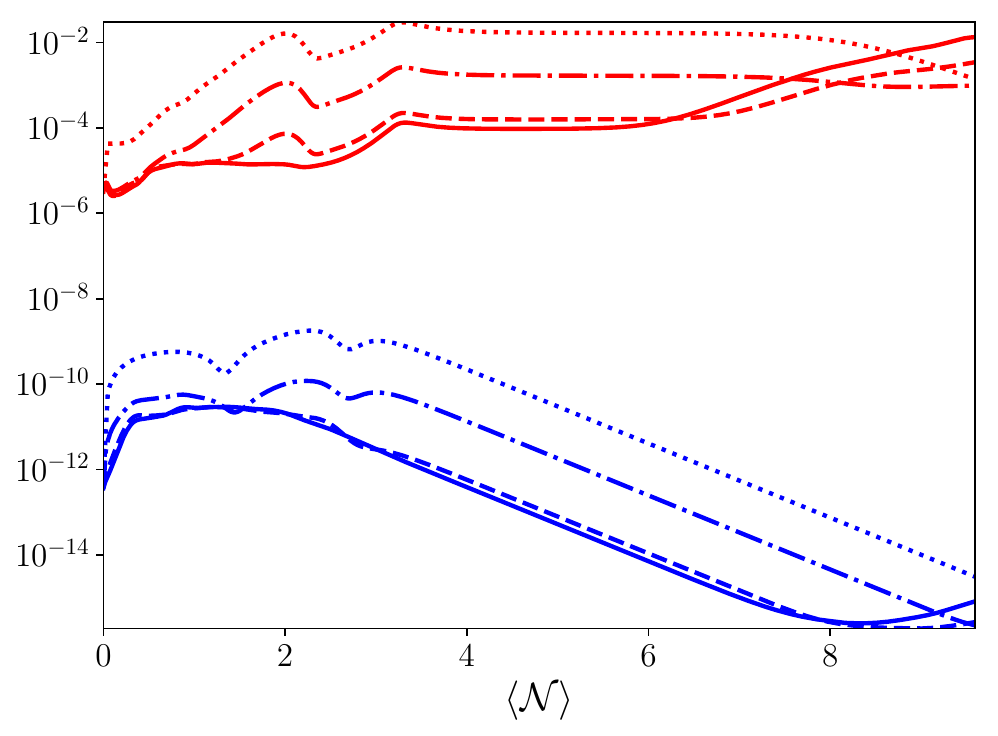}
\caption{Inflection inflation}
    \end{subfigure}
\caption{\justifying \new{ Convergence tests for the mean hamiltonian (blue) and momentum (red) relative constraints ($\langle|{\cal C}|/[{\cal C}]\rangle$, ${\cal C }= {\cal H}, {\cal M}$) across multiple runs with $N=32$ (dotted), $N=64$ (dash-dotted), $N=128$ (dashed), and $N = 256$ (solid). $k_{\text{max}}^{\cal R}/dk$ is set to $8$ across all runs, giving about $\ln 8\simeq 2$ efolds of stochastic sourcing. See Appendix \ref{app:conv} for interpretation.}}
\label{fig:Convergence}
\end{figure*}

\new{\textbf{Constraints.} Let us recall the \textit{absolute constraint magnitudes} $[{\cal H}]$ and $[{\cal M}]$, calculated by summing the absolute value of the terms in the BSSN version of the constraint equations \cite{Launay2025}
\begin{equation}
\left\{
    \begin{aligned}
[{\cal H}] & \equiv  \big|{ }^{3}R\big|+\big|\frac{2}{3} K^2\big|+\big|\tilde{A}_{i j} {A}^{i j}\big|+\big|2 M_{Pl}^{-2} \rho\big|, \\
[{\cal M}_i] & \equiv \big|\tilde{\gamma}^{kl}  \partial_k\tilde{A}_{il} + 2\tilde{\gamma}^{kl}\tilde{\Gamma}^m_{l(i}\tilde{A}_{k)m}\big| \\
& +\big| 3\tilde{\gamma}^{kl}\tilde{A}_{ik}\frac{\partial_l\chi}{2\chi}\big| +\big|\frac{2}{3} K_{\mid i}\big|+\big|M_{Pl}^{-2} {\cal { J}}_i\big|,\\
[{\cal M}] & \equiv \sqrt{\sum_i [{\cal M}_i]^2}.     \end{aligned}\right .
    \label{eq:abscons}
\end{equation}
{These account for the typical magnitude before cancellation. As demonstrated by the lower panels in Figures \ref{fig:quad_contrast} and \ref{fig:usr_contrast}, physicality of the simulations requires $|{\cal H}|/[{\cal H}]\ll 1$ and $|{\cal H}|/[{\cal H}]\ll 1$.} More quantitatively in the case of perturbative inflation, $[{\cal H}]$ is of order $H^2$ from the background, and $[{\cal M}]$ contains linear order gradients. Satisfying the constraints when our self-consistent (linear order) stochastic noise acts therefore means that $|{\cal H}|/[{\cal H}]$ should be of second order and $|{\cal M}|/[{\cal M}]$ first order.

The convergence of the constraints looks at the robustness of the numerical scheme: increasing resolution for the same physics should yield converging accuracy. In Figure \ref{fig:Convergence}, we show that stochastic convergence is attained as we increase $N$: as the sources are first order in the noise, each stochastic kick will violate the constraints at second order as reflected by the amplitudes' early plateau. Once the modes have all entered, the usual inflation-driven exponential decay of the constraints is observed. 

The convergence test loses its meaning later in the evolution as it unfairly disfavors higher resolution runs: either a numerical accuracy floor is reached faster (left) or physics has more small scales to enhance  unphysical noise (right) (see \cite{Launay2025} for the spectral study of the constraints during ultra slow-roll). This does not impact the validity of our simulations as they still satisfy the constraints but rather makes the design of a meaningful and fair convergence test more difficult.

\textbf{Spectra.} A convergence test can also focus on the extraction of quantities of interest and complete the verification where the constraint convergence is blinded. We provide here the convergence of the extracted spectra ${\cal R}^{\rm NR}$ to the linear evolution in Figure \ref{fig:specConvergence}. The extreme linearity of quadratic inflation makes the Mukhanov-Sasaki solution the reference that a higher resolution should approach; indeed is what we find. }

\section{\new{A word on stochastic backreaction} \label{app:stochback}}
\begin{figure}[!ht]
\includegraphics[width=\linewidth]{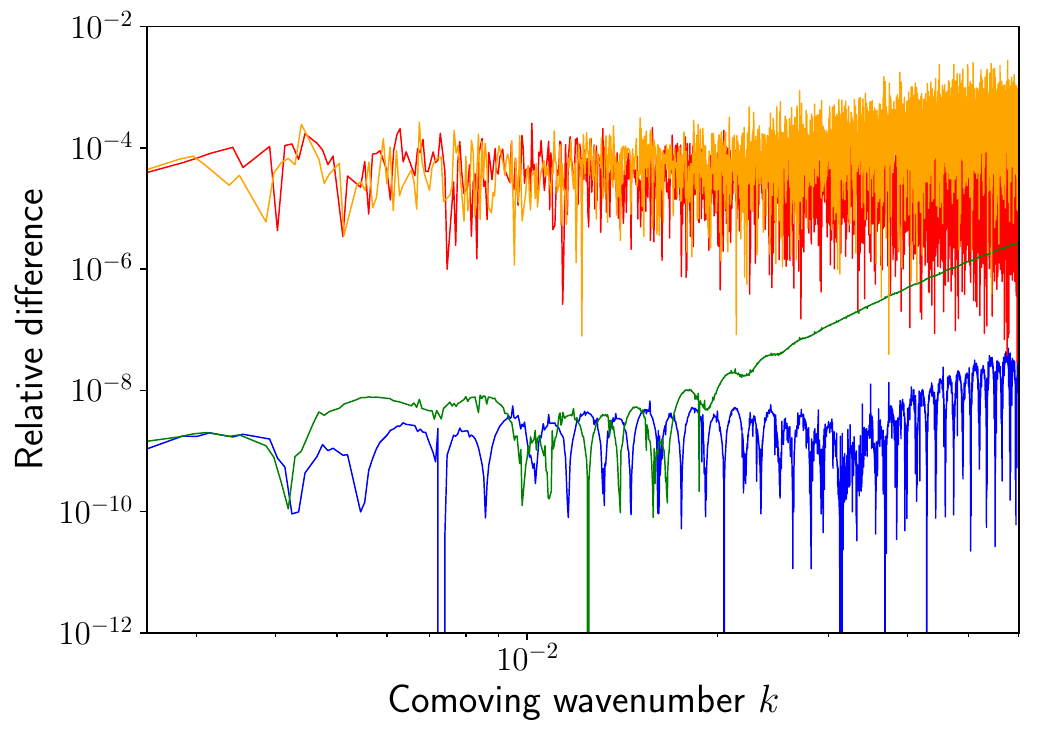}
\caption{\justifying \new{Relative difference in dimensionless ${\cal R}^{\rm NR}$ power spectra between \textit{level} $2$ and \textit{level} $0$ stochastic backreaction (top 2 curves), or between \textit{level} $1$ and \textit{level} $0$ stochastic backreaction (bottom curves), for quadratic inflation (red, blue) and inflection inflation (orange, green). All the same $N=256$ stochastic seed.}}
\label{fig:ablationback}
\end{figure}

\new{Accounting for stochastic backreaction is still an open problem and thus relies on heuristics in the literature. The current implementation of \textsc{ISTORIz} allows three levels of stochastic backreaction. The coefficients appearing in the Mukhanov-Sasaki equation \eqref{eq:ReqISTORIZ1} can be computed from: 
\begin{itemize}
    \item \textit{level} $0$ -- The Friedmann equations, evolved separately with the same initial conditions and numerical scheme.
    \item \textit{level} $1$ -- Their average values over the stochastically evolving IR lattice. 
    \item \textit{level} $2$ --  From the local value of the derived field on the IR lattice.
\end{itemize}

\textit{Level 0} constitutes the original and standard formulation of Stochastic Inflation. In a perturbative regime, any backreaction corrections would be of second order in perturbation theory and so do not matter if the leading behavior is of interest but would be important for e.g. non-gaussianity. Figure \ref{fig:ablationback} compares those corrections in our simulations at a spectral level, indicating the relative importance of level 2 backreaction comparativly to level 0, giving a small but detectable effect on the power spectrum.}

\end{document}